\documentstyle[preprint,aps]{revtex}
\tighten
\begin{document}
\draft
\preprint{Preprint Numbers: \parbox[t]{45mm}{ANL-PHY-8421-TH-96\\ KSUCNR-03-96\\
        nucl-th/9605027}}

\title{Ground-state spectrum of light-quark mesons}

\author{C. J. Burden\footnotemark[1], 
Lu Qian\footnotemark[2], 
C. D. Roberts\footnotemark[3],
P. C. Tandy\footnotemark[2] and
M. J. Thomson\footnotemark[4]\vspace*{0.2\baselineskip}} 
\address{
\footnotemark[1]Department of Theoretical Physics, Research School of Physical
Sciences and Engineering, Australian National University, Canberra ACT 0200,
Australia\vspace*{0.2\baselineskip}\\
\footnotemark[2]Centre for Nuclear Research, Department of Physics, Kent State
University, Kent OH 44242\vspace*{0.2\baselineskip}\\
\footnotemark[3]Physics Division, Bldg. 203, Argonne National Laboratory,
Argonne IL 60439-4843\vspace*{0.2\baselineskip}\\
\footnotemark[4]School of Physics, University of Melbourne, Parkville VIC
3052, Australia\vspace*{0.2\baselineskip} }
\date{15/May/96}
\maketitle
\begin{abstract}
A confining, Goldstone theorem preserving, separable Ansatz for the ladder
kernel of the two-body Bethe-Salpeter equation is constructed from
phenomenologically efficacious $u$, $d$ and $s$ dressed-quark propagators.
The simplicity of the approach is its merit.  It provides a good description
of the ground-state isovector-pseudoscalar, vector and axial-vector meson
spectrum; facilitates an exploration of the relative importance of various
components of the two-body Bethe-Salpeter amplitudes, showing that sub-leading
Dirac components are quantitatively important in the isovector-pseudoscalar
meson channels; and allows a scrutiny of the domain of applicability of
ladder truncation studies.  A colour-antitriplet diquark spectrum is
obtained.  Shortcomings of separable Ans\"atze and the ladder kernel are
highlighted.
\end{abstract}
%
\section{Introduction}
The spectroscopy of light-quark mesons is made interesting because of the
role played by dynamical chiral symmetry breaking, the natural scale of which
is commensurate with other scales in this sector.  It also explores quark and
gluon confinement because most vector and axial-vector mesons have masses
that are more than twice as large as typical constituent-quark masses.
Covariant, constituent-quark potential models are a useful tool in the study
of this problem~\cite{SSV94GM94}.

A salient feature of the strong interaction spectrum is the fact that
$m_\rho^2-m_\pi^2 \approx 30 m_\pi^2$, which may be compared with the
vector-pseudovector splitting $m_{a_1}^2-m_\rho^2 \approx 1.7 m_\rho^2$.
Furthermore, the pion mass must vanish in the chiral limit; i.e., when the
current-quark mass vanishes, whereas $m_{a_1}^2\to m_\rho^2$.  (We note that
a vanishing current-quark mass does not entail a vanishing of the
constituent-quark mass.)  These observations are an indication that the
Goldstone-boson character of the pion is a particular and crucial feature of
the strong-interaction spectrum.  That these features are difficult to
capture in potential models is well illustrated in Refs.~\cite{SSV94GM94}

An efficacious framework for studying meson spectroscopy is provided by the
QCD Dyson-Schwinger equations [DSEs]~\cite{DSErev}, which include the ``QCD
gap-equation'' (quark DSE), that has proven useful in the study of quark
confinement and dynamical chiral symmetry breaking, and the covariant,
two-body bound-state Bethe-Salpeter equations [BSEs].  With one exception,
Ref.~\cite{BRvS96}, all spectroscopic studies to date have employed the
rainbow-ladder truncation of the quark DSE and two-body BSE, which is defined
as follows.  Rainbow approximation specifies that the dressed quark-gluon
vertex in the quark DSE is replaced by the bare vertex:
$\Gamma_\mu^a(k,p)\equiv \gamma_\mu \lambda^a/2$, where
$\{\lambda^a\}_{a=1}^8$ are the colour Gell-Mann matrices.  This equation is
then solved with a given model form of the dressed-gluon propagator,
$D_{\mu\nu}(k)$, to yield a dressed-quark propagator, $S(p)=1/[i\gamma\cdot p
A(p^2)+B(p^2)]$.  The kernel of the ladder-approximation to the two-body BSE
is then the customary ladder kernel but with $D_{\mu\nu}(k)$ and $S(p)$
employed in place of free-particle propagators.  

It has been shown~\cite{DS79} that for any $D_{\mu\nu}(k)$ that leads to the
dynamical generation of a fermion mass in the chiral limit; i.e., to
dynamical chiral symmetry breaking, the isovector-pseudoscalar meson BSE
necessarily admits a $P^2=0$ bound-state solution ($P_\mu$ is the
total-momentum of the dressed-quark, antiquark system).  No fine-tuning is
necessary to ensure this outcome and one thus has a natural understanding of
the pion as both a Goldstone boson and a bound-state of a strongly-dressed
quark and antiquark.  This outcome is the result of an equivalence, in the
chiral limit, between the quark DSE and the isovector, pseudoscalar meson
BSE.  This equivalence is an intrinsic feature of the DSEs, which persists in
more sophisticated truncation schemes~\cite{BRvS96,HJM95}.

The most extensive and phenomenologically successful spectroscopic studies in
the rainbow-ladder framework are those of Ref.~\cite{JM93}, in which the
quark DSE is solved numerically for spacelike-$p^2$ using a model gluon
propagator.  In Landau gauge the behaviour of the gluon propagator is
constrained by perturbation theory for $k^2>1-2$~GeV$^2$~\cite{BP89} and one
models the infrared behaviour, which is presently unknown.  Such studies have
the ability to unify many observables via the few parameters that
characterise the behaviour of the model dressed-gluon propagator in the
infrared.

In this approach, solving the meson BSEs is complicated by the fact that
these equations sample the dressed-quark propagator off the spacelike-$p^2$
axis.  In Ref.~\cite{JM93} this difficulty was circumvented by employing a
derivative expansion of the dressed-quark propagator functions, $A(p^2)$ and
$B(p^2)$, and estimating the error introduced thereby.  This, however,
obscures the discussion and exploration of the role of quark and gluon
confinement, a sufficient condition for which is the absence of a Lehmann
representation for the dressed-quark and dressed-gluon propagators.  The
problem becomes more acute in studies of scattering observables.

An algebraic parametrisation of a confining dressed-quark propagator, based
on numerical solutions of model quark-DSEs, has been used successfully in
studies of a large range of mesonic scattering observables; for example:
$f_\pi$, $r_\pi$, the $\pi$-$\pi$ scattering-lengths and the pion
electromagnetic form factor~\cite{R94}; $f_K$ and the charged and neutral
kaon electromagnetic form factors~\cite{BRT96}; the anomalous
$\gamma\pi\to\gamma$~\cite{FMRT95} and $\gamma\pi\to\pi\pi$ transition form
factors\cite{AR96}.  In these studies the dressed-gluon propagator is only
specified implicitly insofar as the model dressed-quark propagator can be
used as a constraint on its form via the quark DSE.  

It would be useful to make this connection explicit.  However, given a
dressed-quark propagator it is not possible, in principle, to invert the
quark DSE and extract a dressed-gluon propagator; one reason being that the
quark DSE involves the dressed-quark-gluon vertex, which depends implicitly
on both the dressed-quark and dressed-gluon propagators.  In the peculiar
case of the rainbow truncation this particular difficulty, at least, is
eliminated.

It is known that the rainbow truncation is only quantitatively and
qualitatively reliable in Landau gauge~\cite{AtkinetalBP94}, which means that
it is inappropriate to infer a connection between a given
phenomenologically-constrained model dressed-gluon propagator and a solution
of the gluon DSE, such as those obtained in Refs.~\cite{BP89,BBZAtkin83}, in
any other covariant gauge.  The quark DSE is, in general, a pair of coupled,
non-linear integral equations for $A(p^2)$ and $B(p^2)$. In Landau gauge, the
kernel in the equation for $A(p^2)$ is sufficiently complicated, even in
rainbow truncation, that it is not possible to invert the equation without
introducing kinematic singularities.  An explicit connection between the
dressed-quark propagator and a confining dressed-gluon propagator via the
inversion of the quark DSE is therefore not possible.

A goal of this study, and another~\cite{CG95}, is to explore the extent to
which pion and kaon observables, as embodied in the model dressed-quark
propagators employed in Refs.~\cite{R94,BRT96,FMRT95,AR96}, constrain the
properties of all light-quark mesons.  As we have described, it is not
possible to explicitly construct a dressed-gluon propagator from these
dressed-quark propagators.  However, one may adopt a purely phenomenological
approach in order to construct a simple, confining Ansatz for the {\it
kernel} of the two-body BSEs that is constrained by the pion and kaon
scattering observables.  This facilitates the present exploration of the 
extent to which a confining, Goldstone theorem preserving, BSE approach can 
generate the ground state spectrum of light-quark mesons. 
This approach  allows one to easily identify those
channels to which it is applicable and those in which it is inadequate, and
to explore the extent to which observable properties are influenced by
sub-leading Dirac components in the Bethe-Salpeter amplitude; e.g., the
influence of the pseudovector, $\gamma_5\gamma\cdot P$, piece of the pion
Bethe-Salpeter amplitude on the pion mass and decay constant.  Such terms
have been neglected in almost all studies undertaken to date.   A similar
situation holds for the calculation of hadronic  coupling constants and
associated form factors for processes such as 
$\rho \rightarrow \pi\pi$~\cite{HRM92,MT} and 
$\rho \rightarrow \gamma\pi$~\cite{T96}.  Semi-phenomenological $\bar{q}q$ 
Bethe-Salpeter amplitudes for the leading Dirac covariant are currently used 
to facilitate the necessary integrations.  The amplitudes provided here have 
a significant amount of dynamical justification and yet are simple enough to
allow a more realistic study of hadron couplings.   We also take this
opportunity to explore diquark correlations.  

It is with this goal in mind that, in Sec.~\ref{sectwo}, we construct a
constrained, confining, flavour-dependent, separable Ansatz for the
dressed-ladder kernel of the two-body BSEs.  Light-quark meson spectroscopy
is discussed in Sec.~\ref{secthree}.  In Sec.~\ref{secfour} we consider the
spectroscopy of colour-antitriplet quark-quark (diquark) correlations, which
are bound in dressed-ladder truncation.  This is a defect of the truncation,
which is due to the fact that there are no repulsive terms in the kernel at
this level of truncation.  It is a peculiarity; repulsive terms appear at
every higher order, with ``order'' referring to the number of explicit
dressed-gluon propagators in the kernel, and these eliminate the
diquark bound states~\cite{BRvS96}.  Diquark spectroscopy is nevertheless of
contemporary interest because a number of studies of the nucleon Fadde'ev
equation have proceeded under the assumption that the quark-quark 
${\cal T}$-matrix can be represented as a sum of simple diquark-pole terms, 
and that the
mass splittings are such that only the lowest mass poles need be retained in
solving the reduced two-body equation that results~\cite{C92,HT94}.  We
summarise and conclude in Sec.~\ref{secfive}.

\section{Separable Ansatz for the Bethe-Salpeter Kernel}
\label{sectwo}
The Dyson-Schwinger equation for the Euclidean-space dressed-quark propagator
(Schwinger function),
\begin{equation}
\label{Sform}
S(p) = -i\gamma\cdot p \, \sigma_V(p^2) + \sigma_S(p^2)
= \frac{1}{i\gamma\cdot p\, A(p^2) + B(p^2)}~,
\end{equation}
can be written as
\begin{equation}
\label{dseq}
S^{-1}(p) = i \gamma\cdot p + m + 
\int\frac{d^4k}{(2\pi)^4}\,g^2
D_{\mu\nu}((p-k)^2)\,\gamma_\mu\,\frac{\lambda^a}{2}\,S(k)\,\Gamma_\nu^a(k,p)
\end{equation}
where $m$ is the (bare) current-quark mass.  The Euclidean Dirac matrices
satisfy the algebra $\{\gamma_\mu,\gamma_\nu\}=2\delta_{\mu\nu}$, where
$\delta_{\mu\nu}$ is the Kronecker-delta, and \mbox{$a\cdot b \equiv
\sum_{i=1}^4\,a_i b_i$}.  In Eq.~(\ref{dseq}), $D_{\mu\nu}(k)$ is the
dressed-gluon propagator and $\Gamma_\mu(k,p)$ is the dressed quark-gluon 
vertex.

The homogeneous BSE for a quark-antiquark bound state is
\begin{equation}
\label{bsem}
\Gamma^{rs}(k;P) =  \int\frac{d^4q}{(2\pi)^4}\,
K^{rs;tu}(q,p;P)\left[S_{f_1}(q+\xi P)\Gamma(q;P)
	S_{f_2}(q-(1-\xi)P)\right]^{tu}~,
\end{equation}
where $P$ is the centre-of-mass momentum; the $f_1$-flavour-quark carries
momentum $p_{f_1}= q+\xi P$ and the $\bar f_2$-flavour-antiquark momentum is 
$p_{\bar f_2}= -q+(1-\xi)P$; and $K(q,p;P)$ is the quark-antiquark scattering
kernel.  

The ladder approximation is defined by:
\begin{equation}
\label{giaknl}
K^{rs;tu}(q,p;P) \equiv -g^2\,D_{\mu\nu}(p-q)\,
I_{r_F t_F}\,\left(\frac{\lambda^a}{2}\right)_{r_C t_C}\,\gamma_{r_D t_D}\,
I_{u_F s_F}\left(\frac{\lambda^a}{2}\right)_{u_C s_C}\,\gamma_{u_D s_D}~,
\end{equation}
where $\{F,C,D\}$ indicate flavour, colour and Dirac indices.

In rainbow approximation; i.e., using 
\begin{equation}
\label{rainbow}
\Gamma_\mu^a(k,p) = \gamma_\mu\,\frac{\lambda^a}{2}
\end{equation}
in Eq.~(\ref{dseq}), then, if the dressed-quark propagator is known,
Eq.~(\ref{dseq}) can be used to constrain an Ansatz for the ladder
approximation to the kernel of Eq.~(\ref{bsem}).

Herein, solely because of its inherent simplicity, we employ a model of the
form
\begin{equation}
\label{ggdf}
g^2 D_{\mu\nu}(p-k) = \delta_{\mu\nu}\Delta(p-k)~.
\end{equation}
In being proportional to $\delta_{\mu\nu}$, this has the {\it appearance} of
a Feynman gauge propagator.  The appearance is misleading, however.  A
fundamental Slavnov-Taylor identity in QCD entails that the longitudinal
piece of the dressed-gluon propagator must be independent of interactions.
Equation~(\ref{ggdf}) has the transverse and longitudinal components dressed
in exactly the same way.  A propagator of this form could only arise if the
gauge parameter was chosen so as to completely cancel the transverse
interaction contributions; i.e., if the gauge parameter dependent,
longitudinal piece of the gluon propagator is interaction dependent.
Therefore Eq.~(\ref{ggdf}) can only provide a model effective-potential:
$\Delta(p-k)$, defined in this way, cannot in principle be related to
solutions obtained in studies of the gluon DSE, such as
Refs.~\cite{BP89,BBZAtkin83}.  We will describe Eq.~(\ref{ggdf}) as {\it
Feynman-like} gauge.

One can write, without loss of generality, 
\begin{equation}
\label{dfeynman}
\Delta(p-k) = \sum_{n=0}^\infty\,\Delta_n(p^2,k^2)\,p^n\,k^n\,\case{1}{2^n}\,
U_n(\hat p\cdot\hat k)~,
\end{equation}
where $\hat p$ is the unit-magnitude direction-vector for $p$ and  
$\{U_n(x)\}_{n=0}^\infty$ are the complete set of orthonormal Tschebyshev
functions, which satisfy
\begin{equation}
\frac{2}{\pi}\int_{-1}^1\,dx\,\sqrt{1-x^2}\,U_i(x)\,U_j(x) = \delta_{ij}~.
\end{equation}
Translational invariance is preserved if all contributing Tschebyshev moments
are retained.  

The quark DSE, Eq.~(\ref{dseq}), represents two coupled, non-linear integral
equations for $A(s)$ and $B(s)$, where $s=p^2$.  In rainbow approximation,
Eq.~(\ref{rainbow}), and using Eq.~(\ref{ggdf}), these equations are
\begin{eqnarray}
p^2\,A(p^2) & = &  p^2 +
\frac{8}{3}\int\frac{d^4k}{(2\pi)^4}\,
\Delta((p-k)^2)\,p\cdot k\,\frac{A(k^2)}{k^2 A(k^2)^2 + B(k^2)^2}~,\\
B(p^2) & = & m + \frac{16}{3}\int\frac{d^4k}{(2\pi)^4}\,
\Delta((p-k)^2)\,\frac{B(k^2)}{k^2 A(k^2)^2 + B(k^2)^2}~.
\end{eqnarray}
The simplicity inherent in Feynman-like gauge is obvious.  

Introducing the Tschebyshev expansion for $\Delta(p-k)$, Eq.~(\ref{dfeynman}), 
these equations become
\begin{eqnarray}
\label{aeqn}
A(s) & = & 1
+\frac{1}{24\pi^2}\int_0^\infty\,dt\,t^2\,\Delta_1(s,t)\,\sigma_V(t)~,\\  
\label{beqn}
B(s) & = & m+
\frac{1}{3\pi^2}\int_0^\infty\,dt\,t\,\Delta_0(s,t)\,\sigma_S(t)~,
\end{eqnarray} 
from which one observes that, in rainbow approximation and in Feynman-like
gauge, the quark DSE is only sensitive to the zeroth and first Tschebyshev
moments of $\Delta(p-k)$.  Hence, translational invariance of the kernel of the
quark DSE is not lost as long as the zeroth and first Tschebyshev moments are
retained. 

A constrained kernel can now be obtained by employing a rank-N, separable
approximation for the Tschebyshev moments:
\begin{equation}
\label{sepN}
\Delta_n(s,t) = \sum_{i=1}^N\,F_n^i(s)\,F_n^i(t)~.
\end{equation}
The simplest such approximation is rank-1, which is considered herein; i.e.,
one writes:
\begin{equation}
\label{sepone}
\begin{array}{cc}
F_0^1(s)\equiv G(s) = \frac{1}{b} \left(B(s) - m\right)~, &
F_1^1(s) \equiv F(s) = \frac{1}{a} \left(A(s) -1 \right)~,
\end{array}
\end{equation}
where $a$ and $b$ are fixed constants, which are to be determined, and $A(s)$
and $B(s)$ are the functions that appear in the quark propagator.  As will be
seen below, this particular choice for $F_0^1$ and $F_1^1$ is sufficient to
ensure that Goldstone's theorem is preserved.

Substituting Eqs.~(\ref{sepone}) via Eq.~(\ref{sepN}) into Eqs.~(\ref{aeqn})
and (\ref{beqn}) one finds that this latter pair of equations [i.e., the
quark DSE] is solved if, and only if,
\begin{eqnarray}
\label{aparam}
a^2 & = & \frac{1}{24\pi^2}\int_0^\infty\,dt\,t^2\,[A(t)-1]\,\sigma_V(t)~,\\
\label{bparam}
b^2 & = & \frac{1}{3\pi^2}\int_0^\infty\,dt\,t\,[B(t)-m]\,\sigma_S(t)~.
\end{eqnarray}

One now has a rank-1, separable Ansatz for the kernel of the BSE for
like-quarks, which is completely determined by the propagator of that quark;
i.e., Eq.~(\ref{giaknl}) with
\begin{equation}
\label{sepeqm}
g^2 D_{\mu\nu}(p-k) = \delta_{\mu\nu}\Delta(p-k)
= \delta_{\mu\nu}\left[
G(p^2)\,G(k^2)\,+\,p\cdot k\,F(p^2)\,F(k^2) \right]~.
\end{equation}
With $u$- and $d$-quarks treated as indistinguishable, except for their
electric charge, Eq.~(\ref{sepeqm}) can be used in the study of the BSE for
$\pi$, $\omega$ and $\rho$ mesons, for example.

A simple generalisation of this Ansatz to meson-like bound states with arbitrary
flavour content is obtained via the identification  
\begin{eqnarray}
\label{sepneqm}
\lefteqn{S_{f_1}(k+\xi P)\,\Delta(p-k)\,S_{f_2}(k-(1-\xi) P) \equiv}\\
&& \nonumber
S_{f_1}(k+\xi P)\left\{\case{1}{2}\left[ G_{f_1}(p^2)\,G_{f_2}(k^2)
+ G_{f_1}(k^2)\,G_{f_2}(p^2)\right] \right.\\
&& \nonumber \left. + p\cdot k \,\case{1}{2}
\left[ F_{f_1}(p^2)\,F_{f_2}(k^2) + F_{f_1}(k^2)\,F_{f_2}(p^2)\right]\right\}
S_{f_2}(k-(1-\xi) P)
\end{eqnarray}
wherever it appears in the kernel of a given BSE.  This can be used in the
study of the BSE for $K$ and $K^\ast$ mesons, for example.

We observe that, once the propagators for quarks of flavours $f_1$ and $f_2$
are known, Eq.~(\ref{sepneqm}) provides a constrained, separable Ansatz for
the {\em kernel} of the Bethe-Salpeter equation.  If the dressed-quark
propagators have no Lehmann representation then this kernel is free of quark
and gluon production thresholds and may therefore be described as confining.
As remarked above, this Ansatz for the kernel {\em is not} equivalent to an
Ansatz for the gluon propagator and it is inappropriate to infer comparisons
with solutions obtained in studies of the gluon DSE, such as
Refs.~\cite{BP89,BBZAtkin83}.  Such comparisons can only be made when one
employs a gauge-fixing procedure that does not violate the relevant
Slavnov-Taylor identity; for example, Ref.~\cite{FR96}, which employs Landau
gauge and is not separable.  We note that any attempt to construct a
constrained, separable Ansatz in other than Feynman-like gauge will introduce
kinematic singularities in the analogue of Eq.~(\ref{sepneqm}).

The BSE is solved in the rest frame by setting $P=(0,0,0,iM)$ in
Eq.~(\ref{bsem}) and details of this for the case of identical quarks ($f_1 =
f_2$) are given in Appendix~\ref{appA}.  The generalisation to meson-like
bound states with arbitrary flavour content is straightforward using
Eq.~(\ref{sepneqm}).  In general, the ladder truncation of the BSE reduces to
a finite matrix equation that admits solutions for discrete values of the
meson mass, $M$.

\subsection{Meson Decay Constant}
The canonical normalisation of the Bethe-Salpeter amplitude $\Gamma$ is given by
\cite{LS68}
\begin{eqnarray}
\label{gnorm}
2  P_{\mu} & =& N_c \int \frac{d^4k}{(2\pi)^4} \left\{ \rule{0mm}{7mm}
{\rm tr}_{D} \left[ \overline{\Gamma}(k,-P) \partial_\mu^P S_{f_1}(k + \xi P)
\Gamma(k,P) S_{f_2}(k - (1 - \xi)P)\right]  \right. \\
 & & \nonumber    \left.
 + {\rm tr}_{D}\left[ \overline{\Gamma}(k,-P) S_{f_1}(k + \xi P)\Gamma(k,P)
   \partial_\mu^P S_{f_2}(k - (1 - \xi)P) \right] \rule{0mm}{7mm}\right\},
\end{eqnarray}
where $\overline{\Gamma}(k,P)^{\rm T} = C^{-1}\Gamma(-k,P)C$ defines the
corresponding anti-meson amplitude.  In ladder approximation the kernel of
the Bethe-Salpeter equation is independent of the centre-of-mass momentum,
$P$, hence there is no contribution of the type $\partial K/\partial P$ to
the normalisation.

The pseudoscalar meson decay constant, $f_{\cal P}$, is defined by:
\begin{equation}
\langle 0| \overline{\Psi}(0) \gamma_\mu \gamma_5
     \frac{\Lambda^{\cal P}}{2}\Psi(0) | \Phi(P) \rangle
	 = P_\mu f_{\cal P},
\end{equation}
where $|\Phi(P) \rangle$ is the pseudoscalar meson state vector, 
$\Lambda^{\cal P}$ are matrices acting in flavour space and $\Psi$ is a colour
triplet and flavour multiplet of Dirac spinors.  For the $K^-$ meson, for
example, the relevant flavour matrix is, with $\{\lambda_i\}_{i=1}^8$ the
Gell-Mann matrices,
\begin{equation}
\Lambda^{K^-} = \frac{1}{\surd 2}(\lambda_4 + i\lambda_5) =
    \left( \begin{array}{ccc}
     0 & 0 & \surd 2 \\ 0 & 0 & 0 \\ 0 & 0 & 0  \end{array} \right)~,
\end{equation}
which gives
\mbox{$
 \langle 0| \overline{\Psi}_{u}(0) \gamma_\mu \gamma_5
     \Psi_{s}(0) | \Phi_{K^-}(P) \rangle  = \surd 2\,P_\mu f_{K^-}~. 
$}  
Thus, the decay constants for the pseudoscalar meson solutions to the BSE
given in Eq.~(\ref{bsem}) are defined by
\begin{equation}
\surd 2\,P_\mu f_{M} = \langle 0| \overline{\Psi}_{f_2}(0) \gamma_\mu \gamma_5
     \Psi_{f_1}(0) | \Phi_{M}(P) \rangle ~. 
\label{fdef}
\end{equation}

To obtain an expression in terms of the Bethe-Salpeter amplitude,
we note that the unamputated BS wave-function  
\begin{equation}
\chi(p,P) = S_{f_1}(p + \xi P) \Gamma(p,P) S_{f_2}(p - (1-\xi)P) ,
\end{equation}
can be expressed as
\begin{eqnarray}
\lefteqn{(2\pi)^4 \delta^4(p-q) \chi(p,P) }\nonumber \\
 & = & \int d^4x d^4y  \, e^{-iP\cdot [(\xi x + (1-\xi) y] }
    e^{-i(q\cdot x - p\cdot y)}
       \langle 0| \Psi_{f_1}(x)\overline{\Psi}_{f_2}(y) |\Phi(P) \rangle.
\end{eqnarray}
(For a colour singlet bound state, $\chi(p;P)$ is diagonal in colour-space.)
Multiplying both sides by $\gamma_5\gamma\cdot P$, taking the matrix trace
throughout, evaluating the integrals over $p$ and $q$ and using
Eq.~(\ref{fdef}) one obtains
\begin{equation}
P^2 f_M = \frac{N_c}{\surd 2} \int \frac{d^4p}{(2\pi)^4} {\rm tr}_{D} 
\left[\gamma_5\gamma\cdot P 
	S_{f_1}(p + \xi P) \Gamma(p,P) S_{f_2}(p + (1-\xi)P)\right]~,
\label{decay}                                              
\end{equation}
which provides the relation between the Bethe-Salpeter amplitude and the
canonically defined meson decay constant.  In this equation $\Gamma$ is
normalised according to Eq.~(\ref{gnorm}). 

\subsection{Dressed Quark Propagators.}
The separable Ansatz is completely defined once the quark propagators are
specified.  Following Ref.~\cite{BRT96}, the scalar and vector parts of the 
quark propagators are defined in terms of dimensionless functions: 
\begin{equation}
\sigma_V^f(s) = \frac{1}{2D}\bar\sigma_V^f\left(x\right), \hspace{5 mm}
\sigma_S^f(s) = \frac{1}{\sqrt{2D}}\bar\sigma_S^f\left(x\right)~, 
\end{equation}
with $s=p^2$, $x=s/(2 D)$, $D$ is a mass-scale parameter, and where
($\Lambda=10^{-4}$): 
\begin{eqnarray}
\lefteqn{\bar\sigma_S^f(x)  =  
  \frac{\overline{m}_f}{x + \overline{m}_f^2}
      \left(1 - e^{-2(x + \overline{m}_f^2)}\right) + }\nonumber\\
& &  \frac{1 - e^{-b_1^f x}}{b_1^f x} \frac{1 -
e^{-b_3^fx}}{b_3^fx}
    \left(b_0^f + b_2^f  \frac{1 - e^{-\Lambda x}}{\Lambda x}\right) ,
\label{ssb}
\end{eqnarray}
and
\begin{eqnarray}
\bar\sigma_V^f(x) = \frac{2(x + \overline{m}_f^2)
   - 1 + e^{-2(x + \overline{m}_f^2)}}{2(x + \overline{m}_f^2)^2}.
\label{svb}
\end{eqnarray}
Here \mbox{$\overline{m}_f = m_f/ \sqrt{2D}$}. In this work
the $u$ and $d$ quarks are considered to be identical, except for their
electric charge.

The dressed-quark propagator described by Eqs.~(\ref{ssb}) and (\ref{svb}) is
an entire function in the finite complex $p^2$-plane and may therefore be
interpreted as describing a confined particle\cite{DSErev}. The $\sim{\rm
e^{-x}}$ form that ensures this is suggested by the algebraic solution of the
model DSE studied in Ref.~\cite{BRW92}, which employed a confining model
dressed-gluon propagator and dressed quark-gluon vertex.  Furthermore, the
behaviour of Eqs.~(\ref{ssb}) and (\ref{svb}) on the spacelike-$p^2$ axis is
such that, neglecting $\ln[p^2]$ corrections associated with the anomalous
dimension of the dressed-quark propagator in QCD, which are quantitatively
unimportant herein, asymptotic freedom is manifest.  In Eq.~(\ref{ssb}) the
term $\sim 1/x^2$ allows for the representation of dynamical chiral symmetry
breaking and the $\sim m/x$ term represents explicit chiral symmetry
breaking.
 
In Ref.~\cite{BRT96} the five parameters $\{\bar m_u,b_0^u,\ldots,b_3^u\}$ in
Eqs.~(\ref{ssb}) and (\ref{svb}) were varied in order to determine whether this
model form could provide a good description of the pion observables:
$f_\pi$; $m_\pi$; $\langle\bar q q\rangle$; $r_\pi$; the $\pi$-$\pi$ scattering
lengths and partial wave amplitudes; and the electromagnetic pion form factor. 
A very good fit was found with the $u$-quark parameter values listed in
Eq.~(\ref{tableparam}):
\begin{equation}
\label{tableparam}
\begin{array}{lll}
   &   \mbox{$u$-quark} &   \mbox{$s$-quark} \\
 \bar m_f & 0.00897 & 0.224\\
 b_0^f  &   0.131     & 0.105  \\
 b_1^f  &   2.90     & 2.90      \\
 b_2^f  &   0.603    & 0.740      \\
 b_3^f  &   0.185     & 0.185      
\end{array}
\end{equation}
The scale is set with $D=0.160$~GeV$^2$.  This same model also provides a
good description of the $\gamma^\ast\pi\rightarrow\gamma$~\cite{FMRT95} and
$\gamma\pi^\ast\to\pi\pi$~\cite{AR96} transition form factors.

Dyson-Schwinger equation studies~\cite{WKR91} indicate that while it is a
good approximation to represent the $u$- and $d$-quarks by the same
propagator, this is not true for the $s$-quark.  For example; contemporary
theoretical studies suggest that \mbox{$2m_s/(m_u+m_d)\sim
17-25$}~\cite{PDG94} and \mbox{$\langle\bar s s\rangle\sim 0.5-0.8 \;\langle
\bar u u\rangle$}\cite{Nar}, which is a nonperturbative difference.  In
Ref.~\cite{BRT96}, with this in mind, the model forms in Eqs.~(\ref{ssb}) and
(\ref{svb}) were employed in a study of the kaon observables: $f_K$;
$\langle\bar s s\rangle$; $r_{K^0}$; $r_{K^\pm}$; and the electromagnetic
form factors of the charged and neutral kaon.  The sensitivity of these
observables to $\bar m_s$ and $\langle\bar s s\rangle$ was too weak for an
independent determination and therefore $\bar m_s=25 \bar m_u$ and $b_0^s=0.8
b_0^u$, which ensures $\langle\bar s s\rangle=0.8\langle\bar u u\rangle$,
were chosen for consistency with other theoretical estimates.  The parameter
$b_2^s$ was allowed to vary to provide a minimal residual difference between
the $u/d$- and $s$-quark propagators and a very good fit to the kaon
observables was obtained with the value listed in Eq.~(\ref{tableparam}).

To complete the specification of the constrained separable approximation to the
kernel of  the Bethe-Salpeter equation, the quantities $a$ and $b$ in
Eqs.~(\ref{aparam}) and (\ref{bparam}) must be determined.  However, using
Eqs.~(\ref{ssb}) and (\ref{svb}) neither $a$ nor $b$ is finite. 
Equations~(\ref{aparam}) and (\ref{bparam}) only yield finite values if the 
large
spacelike-$x$ behaviour of $\bar\sigma_V$ and $\bar\sigma_S$ is such that:
\begin{equation}
\begin{array}{cc}
\displaystyle \bar\sigma_V(x) = \frac{1}{x+\bar m^2}
	 + \mbox{O}\left(\frac{1}{x^{2+\delta}}\right) ~,\; & \;
\displaystyle \bar\sigma_S(x) = \frac{\bar m}{x+\bar m^2}
	 + \mbox{O}\left(\frac{1}{x^{2+\delta}}\right)~,
\end{array}
\end{equation}
for any $\delta>0$.  Dynamical chiral symmetry breaking in QCD entails that
at large $x$ [up to corrections $\sim[\ln x]^{-\gamma}$, $\gamma<1$]
$\bar\sigma_S(x) = \bar m/(x+\bar m^2) + {\rm O}(x^{-2})$ and hence no quark
propagator that properly incorporates the momentum dependence at large-$x$
due to dynamical chiral symmetry breaking will yield finite values of $a$ and
$b$.  (This behaviour is tied to the necessary divergence of the quark
condensates in QCD; necessary because condensates are related to two-point
Schwinger functions evaluated at zero relative Euclidean spatial separation.)

To complete the specification of the constrained, separable Ansatz one must
therefore incorporate an ultraviolet regularisation in the propagator:
\begin{eqnarray}
\lefteqn{\bar\sigma_S^{f\,{\rm Reg}}(x)  = 
\frac{\hat{m}_f}{x + \hat{m}_f^2}
      \left(1 - e^{-2(x + \hat{m}_f^2)}\right) + } \nonumber \\
& & 
  \frac{1 - e^{-b_1^fx}}{b_1^fx} \frac{1 - e^{-b_3^fx}}{b_3^fx}
    \left(b_0^f + b_2^f  \frac{1 - e^{-b_4x}}{b_4x}\right)
\frac{1 - e^{-(\epsilon_{\rm S}^f x)^2}}{(\epsilon_{\rm S}^f x)^2}
\label{ssreg}
\end{eqnarray}
\begin{equation}
\label{svreg}
\bar\sigma_V^{f\,{\rm Reg}}(x) = \frac{2(x + \hat{m}_f^2)
   - e^{-\epsilon_{\rm V}^2(x + \hat{m}_f^2)^2}
      + e^{-2(x + \hat{m}_f^2)}}{2(x + \hat{m}_f^2)^2},
\end{equation} 
which introduces three new parameters: $\epsilon_V$, $\epsilon_S^u$,
$\epsilon_S^s$, that are not determined by the studies of Ref.~\cite{BRT96}.
The parameter $\epsilon_V=0.1$ is chosen so as to ensure that
$\bar\sigma_{V}^{f\,{\rm Reg}}$ are numerically good approximations to
$\bar\sigma_{V}^f$ on the domain $0<x<3$; our results are not sensitive to
the domain $x>3$.  It is not varied but we have established that our results
are insensitive to it; i.e., that changes can be absorbed into a change in
$\epsilon_S^f$. The regularisation parameters modify the large-$p^2$
behaviour of the propagator, which entails that the light-quark mass values
must be re-fit ($\bar m_q \rightarrow \hat m_q$).

Equations~(\ref{sepone}-\ref{bparam}), (\ref{sepneqm}), (\ref{tableparam}),
(\ref{ssreg}) and (\ref{svreg}) completely specify the constrained,
confining, separable Ansatz for the ladder-kernel of the Bethe-Salpeter
equation.  The numerical studies proceed by varying the four parameters $\hat
m_f$ and $\epsilon_S^f$ in order to fit $f_{\pi/K}$ and $m_{\pi/K}$ and then
predicting the ground state spectrum of octet mesons.  Diquark systems are
also studied.

\section{Mesons}
\label{secthree}
The Bethe-Salpeter equation considered for a bound state of a quark of
flavour $f_1$ and an antiquark of flavour $\bar f_2$ is
\begin{equation}
\Gamma(p,P) = - \frac{4}{3} \int \frac{d^4q}{(2\pi)^4} 
\Delta(p - q) \gamma_\mu S_{f_1}(q + \xi P) \Gamma(q,P) 
	S_{f_2}(q - (1-\xi) P) \gamma_\mu.  \label{bspi}
\end{equation}
In this equation $\Delta(p-q)$ is obtained from Eq.~(\ref{sepneqm})
with $S_{f_i}$ obtained from Eqs.~(\ref{ssreg}) and (\ref{svreg}) using the
parameters in Eq.~(\ref{tableparam}) and Table~\ref{tabnewparam}.  This equation
is solved in each channel as an eigenvalue problem of the form \mbox{$K\Gamma =
\lambda(P^2)\Gamma$}, with the bound state mass identified from
\mbox{$\lambda(P^2=-M^2)=1$}.

\subsection{Scalar and Pseudoscalar Mesons.}

\subsubsection{$f_1=u/d=f_2$}
In this case the requirement of charge conjugation invariance for the neutral
mesons entails $\xi=1/2$.  The form of the charge parity ${\cal C} = \pm$, 
$f_1=u/d=f_2$ Bethe-Salpeter amplitudes, $\Gamma_{\cal C}$, obtained as 
solutions at the mass-shell point $P^2=-M^2$ are given in
Eqs.~(\ref{gampseud}-\ref{gamscmn}).  

These equations expose a shortcoming of separable Ans\"atze: the pseudoscalar
and pseudovector pieces of the pseudoscalar Bethe-Salpeter amplitude are
characterised by the same function, which is not the case in general.  

The calculated eigen-vectors are given in Eq.~(\ref{uures}) and the bound
state masses in Table~\ref{tabmasses}.

The separable Ansatz for the kernel of the Bethe-Salpeter equation yields
$m_{0^{-+}}=0$ when \mbox{$\bar m_{u/d}=0$}.  This is a necessary consequence
of the equivalence between the isovector-pseudoscalar BSE and the quark DSE
in this chiral limit\cite{DS79}, which is preserved in the approach described
herein and discussed in detail in Ref.~\cite{BRvS96}.

One might be tempted to conclude from Eq.~(\ref{uures}) that for ${\cal C}=+$
states the leading Dirac component of the amplitude dominates; i.e., the pure
$\gamma_5$ component dominates for the $0^{-+}$ state and the pure $I_D$
component for the $0^{++}$ state.  Indeed, it is an often used approximation
to neglect sub-leading Dirac components of the Bethe-Salpeter amplitude in
ground state studies using the Bethe-Salpeter equation.  Considering
Tables~\ref{tabmasses} and \ref{tabwkdcay} one observes that while this is a
good approximation for the heavy $0^{++}$ state, it represents an erroneous
conclusion for the light $0^{-+}$ state, for which the sub-leading,
axial-vector component provides 17\% of the mass and 39\% of the decay
constant.  This feature is also seen in Ref.~\cite{BRvS96}.

Table~\ref{tabmasses} shows that the separable Ansatz for the BSE kernel
yields a large $m_{0^{++}} -m_{0^{-+}}$ splitting without fine tuning, thus
reproducing this characteristic feature of the strong interaction spectrum.
The $0^{++}$ state can be identified with the $a_0(980)$ meson.  The
discrepancy between the calculated and observed masses is consistent with the
contention that this state involves a considerable $\bar K$-$K$ admixture,
which can be represented as a contribution to the Bethe-Salpeter kernel but
is absent in ladder approximation.

No $J^{PC}=0^{+-}$ states have been observed in the strong interaction
spectrum.  However, in general, as observed in Ref.~\cite{LS69}, the
Bethe-Salpeter equation admits solutions of this type.  
The amplitudes for such solutions characteristically differ from their 
${\cal C}=+$ counterparts by the factor $p \cdot P$ which is odd under charge 
conjugation. Such states have no analogue in quantum mechanics
since, for equal-mass constituent-particles on shell, \mbox{$p\cdot P=0$}.  In
this context one observes that our separable Ansatz for the kernel of the
Bethe-Salpeter equation yields very heavy $0^{--}$ and $0^{+-}$ states, with
\mbox{$m_{0^{--}}\sim 10\, m_{0^{-+}}$} and \mbox{$m_{0^{+-}}\sim 2\,
m_{0^{++}}$}; i.e., it yields results consistent with the observed strong
interaction spectrum.

\subsubsection{$f_1=u/d$, $f_2=s$}
\label{usmesons}
The Bethe-Salpeter equation for $\overline{u}$-$s$ states is Eq.~(\ref{bspi})
with $f_1=u/d$, $f_2=s$.  Consider first the pseudoscalar (kaon) channel.  A
value of $\xi$ is determined by ensuring that the electric charge of the
$K^0$ is zero in impulse approximation~\cite{BRT96}.  The value obtained in
Ref.~\cite{BRT96} with empirical Bethe-Salpeter amplitudes is
$\xi=0.49\;(\approx 0.5)$, while the present work requires
$\xi=0.56\;(\approx 0.5)$.  There is only a weak sensitivity of masses to 
changes in $\xi$ of this magnitude throughout this work.

The solution amplitude is given in Eq.~(\ref{ampk}) with the calculated
eigen-vector given in Eq.~(\ref{useigenvec}) and the mass in
Table~\ref{tabmasses}.  We note that there is no charge-parity, ${\cal C}$,
symmetry for bound states of distinguishable quarks.

The leading, pseudoscalar Dirac amplitudes ($\lambda_{1u}$, $\lambda_{1s}$) 
again appear to be dominant for the kaon, however, as for the pion, 
the sub-leading, axial-vector amplitudes ($\lambda_{3u}$, $\lambda_{3s}$) 
contribute significantly to the mass (17\%) and decay constant (33\%).

Each type of covariant in the kaon solution is weighted by two amplitudes
that describe the internal momentum dependence in terms of functions that
relate to the dressed-propagators of the $u/d$- and $s$-quarks.  These are
found to have approximately equal influence in the solution.  For example,
from Eq.~(\ref{useigenvec}) and Table~\ref{tabnewparam} one calculates that
$\lambda_1 /b_u = 9.4$~GeV$^{-1}$ and $\lambda_2 /b_s=9.2$~GeV$^{-1}$.  This
means that, using the constrained, separable Ansatz, the kaon Bethe-Salpeter
amplitude for the pseudoscalar covariant is an approximately even mixture of
the $u$- and $s$-quark mass functions.

It is clear from Table~\ref{tabnewparam} that
\begin{equation}
\label{mssrto}
\frac{2\,\hat m_s}{\hat m_u + \hat m_d} = 24.4~,
\end{equation}
which is essentially the same as the ratio obtained
from the mass values in Eq.~(\ref{tableparam}) and is in the range
($17$-$25$) suggested by other theoretical analyses\cite{PDG94}.  With
$D=0.160$~GeV$^2$, $\hat m_{u/d} = 0.00811$ corresponds to $m_{u/d}=4.6$~MeV
and $\hat m_s = 0.198$ corresponds to $m_s = 112$~MeV.  These values {\it
should not} be compared directly with values of $m_{u/d,s}^{\mu^2 = 1~{\rm
GeV}^2}$ quoted by other authors because the regularisation of the vacuum
condensates employed herein, via the parameters $\epsilon_V$ and
$\epsilon_S^f$ in Eqs.~(\ref{ssreg}) and (\ref{svreg}), is unconventional and
enters through the quantities $a$ and $b$ in Eqs.~(\ref{aparam}) and
(\ref{bparam}).  The ratio, Eq.~(\ref{mssrto}), is likely to be less
sensitive to this difference and therefore provides a meaningful point of
comparison.  

The dressed-quark propagators we employ are confining, with the dressed-quark
mass being a function of $p^2$, $M(p^2)$, such that there is no dressed-quark
mass-pole; i.e., no solution of the equation~$p^2 [A^f(p^2)]^2 + [B^f(p^2)]^2
=0$.  A simple estimate of the value of the mass function that is most important
in calculations of meson observables is obtained from the solution of $- p^2
[A^f(p^2)]^2 + [B^f(p^2)]^2 =0$, which might be called the Euclidean
constituent-quark mass, $M_E^f$.  With the parameter values used herein,
$M_E^u=315$~MeV and $M_E^s=397$~MeV.

Our calculations yield no true $0^+$ eigenstate with a mass less than $2~{\rm
GeV}$, which we consider to be the upper limit for the present approach.  The
condition for an eigenstate was closest to being satisfied at a mass of
$1.18~{\rm GeV}$.  This is again consistent with a large $m_{0^+} - m_{0^-}$
splitting without fine tuning.  This $0^+$ state might be identified with the
$K_0^\ast(1430)$.  Such an identification would suggest that this state, like
the $a_0(980)$, has a sizeable coupling to other channels, which contribute
to its mass; i.e., that the ladder kernel is inadequate to properly describe
this channel.

\subsubsection{$\eta$ Meson.} 
Ladder approximation is inadequate to properly study the $\eta$-$\eta'$
complex.  A minimal extension that can dynamically couple the flavour octet
and singlet channels is the inclusion of timelike-gluon exchange diagrams.
This is not considered here.

Instead we study
\begin{eqnarray}
\lefteqn{\Gamma_\eta(p,P)  =  - \frac{4}{3} \int \frac{d^4q}{(2\pi)^4} } 
	\nonumber \\
& & \! \left[ \case{1}{3} (\cos \theta_P - \surd{2}\sin \theta_P)^2
\Delta_u(p-q)\,\gamma_\mu S_u(q + \case{1}{2} P)
		\Gamma_\eta(q,P) S_u(q - \case{1}{2} P) \gamma_\mu \right. + 
	\nonumber \\
& & \! \left.  \case{1}{3} (\surd{2}\cos \theta_P + \sin \theta_P)^2
\Delta_s(p-q)\,\gamma_\mu S_s(q + \case{1}{2} P)
		\Gamma_\eta(q,P) S_s(q - \case{1}{2} P) \gamma_\mu \right]~,
\label{ebse}
\end{eqnarray}
with
\begin{equation}
\Delta_{f}(p-q) =   G_f(p^2)\,G_f(q^2)\,+\,p\cdot q\,F_f(p^2)\,F_f(q^2)~.
\end{equation}
Eq.~(\ref{ebse}) is the projected Bethe-Salpeter equation for the meson whose 
flavour structure is
\begin{equation}
F_\eta = \lambda^8 \cos\theta_P - \lambda^0 \sin\theta_P~,
\end{equation}
with $\lambda^0 = \sqrt{2/3}\,\mbox{diag}(1,1,1)$ and $\theta_P$ an
octet-singlet mixing angle.  The exact kernel of the Bethe-Salpeter equation
would lead to a prediction for $\theta_P$.

With the kernel considered herein, $\theta_P$ is treated as an external
parameter on which the mass and other properties of the $\eta$-meson depend.
For example, in this case the expressions for the normalisation of the
Bethe-Salpeter amplitude, Eq.~(\ref{gnorm}), and the decay constant,
Eq.~(\ref{decay}), are $\theta_P$-dependent.  The modified forms are given in
Eqs.~(\ref{enorm}) and (\ref{edecay}), respectively.

The form of the positive charge parity solution of Eq.~(\ref{ebse}) is given
in Eq.~(\ref{etagamma}).  The calculated mass, decay constant and
eigen-vector, at a number of values of $\theta_P$, are given in
Eq.~(\ref{etares}).  The experimental values of the mass and decay constant
are given in Table~\ref{tabmasses}.  

In this case the sub-leading Dirac amplitudes contribute $\sim$~14\% to the
mass and $\sim$~26\% to the decay constant.

The constrained separable Ansatz favours a small positive value for the
mixing angle, $\theta_P$.  This can be compared with $\theta_P = -10^\circ$
estimated in Ref.~\cite{PDG94}.  As remarked therein, however, there are
large uncertainties in this value.

The $\eta^\prime$-meson can be studied via the projection of the
Bethe-Salpeter equation orthogonal to that in Eq.~(\ref{ebse}), which is
obtained from this equation under $\theta_P \rightarrow \theta_P- \pi/2$.  As
remarked above, one expects timelike gluon exchange, forbidden in the
flavour octet channel, to be important in this mainly singlet channel.  
The results in
Eq.~(\ref{etares}), which one might compare with the experimental values of
$M_{\eta'}^{\rm expt.}  = 958$~MeV and $f_{\eta'}= 89.1 \pm 5$ or $77.8 \pm
5$~MeV, may be interpreted as a guide to the importance of such contributions
in this channel and emphasise the necessity to go beyond ladder
approximation for the $\eta'$ state.

\subsection{Vector and Axial-vector Mesons}
The ladder approximation to the Bethe-Salpeter equation for
vector and axial vector mesons is
\begin{equation}
\Gamma_{\nu}(p,P) = - \frac{4}{3} \int \frac{d^4q}{(2\pi)^4} \Delta(p - q) 
\gamma_\mu S_{f_1}(q + \xi P)
	    \Gamma_{\nu}(q,P) S_{f_2}(q - (1-\xi) P) \gamma_\mu~;
\label{bsv} 
\end{equation}
which is identical to Eq.~(\ref{bspi}) except that the Bethe-Salpeter
amplitude carries a Lorentz index.  On-shell vector and axial-vector bound
states are transverse:
\begin{equation}
P_\nu\,\Gamma_\nu(p,P)=0~,
\label{vatrans}
\end{equation}
which constrains the general form of the Bethe-Salpeter amplitude.

\subsubsection{$f_1=u/d=f_2$}
The most general form of the vector meson Bethe-Salpeter amplitude when using
the separable Ansatz is given in Eq.~(\ref{vamp}) and that for
the axial-vector meson is given in Eq.~(\ref{avamp}).  

These equations expose another shortcoming of separable Ans\"atze: the vector
and axial-vector meson Bethe-Salpeter amplitudes are characterised by the
same functions as the pseudoscalar mesons, which is not true in general.  In
Ref.~\cite{JM93}, for example, the vector meson amplitudes were found to be
much narrower in momentum space.

The ladder approximation does not distinguish between $I=0$ and $I=1$, hence
the vector channel corresponds to both the $\omega$ and $\rho$ mesons.
Similarly, the axial-vector channel corresponds to the $f_1$ and $a_1$
mesons.

The calculated mass for these states is presented in Table~\ref{tabmasses}
and the eigen-vectors in Eq.~(\ref{vaeigenvec}).  With pion and kaon physics
used to fix the parameters of the quark propagators as described in
Sec.~\ref{sectwo}, these results are predictions.  

The sub-leading Dirac amplitudes contribute very little to the $J=1$ meson
masses.

The relevant experimental value to compare the vector meson with is
\mbox{$M_{\omega}^{expt}=782~{\rm MeV}$}, since it is known that pion loop
dressing will lower the $\rho$-meson mass while having little effect on the
$\omega$-meson~\cite{HRM92}.  A recent study of this effect~\cite{MT} yields
\mbox{$M_{\omega}-M_{\rho}= 21.0~{\rm MeV}$}.

These results in the $u-d$ sector indicate that the $u/d$ quark propagator
parameters, previously set by pion physics, have produced a separable BSE
kernel that captures the dominant physics for the ground state vector and
axial vector channels.  

\subsubsection{$f_1=u/d, f_2=s$}

The general form of the Bethe-Salpeter amplitude for the $u$-$\bar s$ meson,
which corresponds to the \mbox{$J^{P}=1^{-}$} $K^{*+}$-meson, is given in
Eq.~(\ref{ksamp}).   We choose $\xi$ so as to ensure the neutrality of the 
$K^{*0}$ meson, which produces $\xi=0.49 \approx 0.5$.  

The form of the amplitude for \mbox{$J^{P}=1^{+}$}, which is a
nearly equal mix of $K_1(1270)$ and $K_1(1400)$, is simply $\gamma_5$ times
this.   The corresponding choice for $\xi$ yields $\xi=0.50$.  
The calculated masses are listed in Table~\ref{tabmasses} and the
eigen-vectors in Eq.~(\ref{kslam}).  

The sub-leading Dirac amplitudes contribute little to the masses.

\subsubsection{$f_1=s=f_2$}
The \mbox{$J^{PC}=1^{--}$} $\bar{s}s$ state, identified with the
$\phi$-meson, is computed in exactly the same manner as the
$\omega/\rho$-meson except for the replacement of the $u$-quark propagator
with that for the $s$-quark.  The Bethe-Salpeter amplitude for the 
$\phi$ and for the $1^{++}$ state, identified with the $f_1 (1510)$ meson, 
are described in Appendix~\ref{phiapp}.

The calculated masses are listed in Table~\ref{tabmasses} and the
eigen-vectors in Eq.~(\ref{philam}).  The sub-leading Dirac amplitudes are
again unimportant.

\subsubsection{$J=1$ Summary}
We observe that the $J=1$ meson spectrum is satisfactorily reproduced.  These
higher-mass states explore a larger domain in the complex quark-momentum
plane than do the pion and kaon, which are used to constrain the separable
Ansatz for the ladder kernel.  This is an indication that a successful
description of a subset of hadronic observables can translate into a
uniformly good description of a broad range of phenomena, which is a feature
that underlies many applications of this framework and emphasises the utility
of studies such as that of Ref.~\cite{FR96}.

\section{Diquark Correlations}
\label{secfour}
The derivation of the homogeneous Bethe-Salpeter equation from the
inhomogeneous equation for the two-body ${\cal T}$-matrix proceeds under the
assumption that there exists a bound-state pole in the channel under
consideration.  In QCD, one expects that confinement ensures the absence of
such poles in the quark-quark ${\cal T}$-matrix and hence that there are no
solutions to the homogeneous Bethe-Salpeter equation in any
colour-antitriplet quark-quark (diquark) channel.  This is supported by the
studies of Ref.~\cite{BRvS96}, which indicate, however, that one must proceed
beyond ladder approximation to obtain this result.  In ladder approximation
one finds bound-state, diquark solutions.  This is a defect of the
truncation.

Studies of the nucleon as a bound-state of three dressed-quarks using the
covariant Fadde'ev equation have been undertaken\cite{C92}.  The appearance
of the pole in the ladder approximation to the homogeneous, quark-quark
Bethe-Salpeter equation was used therein to simplify the three-body problem;
i.e., to re-express it as an effective two-body, quark-diquark problem.  This
technique can also be said to underly the study of Ref.~\cite{HT94}.
Presently, the only justification for this Ansatz is the simplicity it
introduces into the problem.

Accepting this approach for the present it is then important to identify
those diquark correlations that contribute significantly to a given
three-body bound-state.  As a guide one might assume that those diquarks
whose mass is greater than that of the three-body bound-state under
consideration would contribute little to the three-body ground-state mass.
Such studies of the ``u/d-diquark spectrum'' have been reported in
Refs.~\cite{PCR89,CRP87}.  Herein we extend these studies to $SU_f(3)$.

The ladder approximation to the homogeneous Bethe-Salpeter
equation for a diquark correlation involving quarks of flavour $f_1$ and
$f_2$ is
\begin{eqnarray}
\label{bsqqC}
\lefteqn{\Gamma_{\bar 3}(p,P) =  }\\
&& \nonumber
- \int \frac{d^4q}{(2\pi)^4} \Delta(p - q)
\gamma_\mu \frac{\lambda^a}{2}\,S_{f_1}(q + \xi P)
		\Gamma_{\bar 3}(q,P) \left(S_{f_2}(-q + (1-\xi) P) \right)^T
	\left(\gamma_\mu\frac{\lambda^a}{2}\right)^T~,
\end{eqnarray}
where $T$ denotes matrix transpose.  The study of such correlations is
simplified if one defines
\begin{equation}
\Gamma_{\bar 3}^C(p,P) \equiv \Gamma_{\bar 3}(p,P) \,C
\end{equation}
where $C=\gamma_2\gamma_4$ is the charge conjugation matrix.  It follows from
Eq.~(\ref{bsqqC}) that this auxiliary amplitude satisfies
\begin{equation}
\Gamma_{\bar 3}^C(p,P) =  -\frac{2}{3} \int \frac{d^4q}{(2\pi)^4} \Delta(p - q)
\gamma_\mu\,S_{f_1}(q + \xi P)\, \Gamma_{\bar 3}^C(q,P) \,
	S_{f_2}(q - (1-\xi) P) \gamma_\mu~.
\label{bsqq}
\end{equation}
It is immediately obvious that Eq.~(\ref{bsqq}) is identical to
Eq.~(\ref{bspi}) but for a reduction in the (purely-attractive) coupling
strength: $4/3\rightarrow 2/3$.  This observation in Ref.~\cite{CRP87}
entailed the result that the mass of the scalar-$(u-d)$ diquark is greater
than $m_\pi$; and that of the vector $(u-u)$, $(u-d)$ and $(d-d)$
correlations is greater than the mass of the $a_1(1280)$-meson.  (This result
is true in an arbitrary covariant gauge and independent of the form of the
gluon propagator.  However, it is peculiar to ladder
approximation.  As discussed in Ref.~\cite{BRvS96}, any other truncation of
the kernel of the Bethe-Salpeter equation introduces repulsive terms that
eliminate the diquark pole.)

\subsection{Scalar and Pseudoscalar Diquarks}
\subsubsection{$f_1=u/d=f_2$}
To obtain the $J^P=0^+$ diquark solution of Eq.~(\ref{bsqqC}) one searches
for the $0^-$ auxiliary amplitude solution of Eq.~(\ref{bsqq}).  The latter
can be written in the form
\begin{equation}
\Gamma_{\bar 3}^C (p,P) = G_u(p^2)\,
\left[\lambda_f^{\bar 3}  -
i\,\lambda_W^{\bar 3} \,\gamma\cdot\hat P \right]i\, \gamma_5~,
\end{equation}
which is identical in form to $\Gamma_+^{\rm pseud}$ in Eq.~(\ref{gampseud}).
The $0^-$ pseudoscalar diquark solution is described by an auxiliary
amplitude identical in form to $\Gamma^{\rm scalar}_+$ of Eq.~(\ref{gamscpl}).
The calculated masses are listed in Table~\ref{massdq} and the eigen-vectors
are given in Eq.~(\ref{dqud}).

One observes that the sub-leading Dirac amplitude contributes 11\% to the
$0^+$ diquark mass.

This result suggests that the $0^+$ diquark-pole will provide a contribution
to the truncated quark-quark ${\cal T}$-matrix that is important in the type
of Fadde'ev equation studies of the nucleon described above.  The much larger
mass found to be associated with the $0^-$ diquark correlation suggests that
it may be neglected in such studies.

\subsubsection{$f_1=u/d$, $f_2=s$}
The homogeneous Bethe-Salpeter equation in the $u/d$-$s$ quark-quark channel
can be written in the form:
\begin{eqnarray}
\Gamma_{us}^{\bar 3\,C}(p,P) =
 & & - \frac{2}{3} \int \frac{d^4q}{(2\pi)^4} 
\Delta(p - q) \gamma_\mu S_u(q + \xi P)
		\Gamma_{us}^{\bar 3\,C}(q,P) S_s(q - (1 - \xi) P) \gamma_\mu, 
\label{bsus}
\end{eqnarray}
where the momentum partitioning parameter is $\xi = 0.56\approx 0.5$, as for
the kaon.  

The solution of this equation that corresponds to the $0^+$ diquark is
identical in form to $\Gamma^{\rm pseud}$ in Eq.~(\ref{ampk}).  The
calculated mass is listed in Table~\ref{massdq} and the eigen-vector in
Eq.~(\ref{usdqev}). 
The sub-leading Dirac amplitudes contribute 11\% to the $0^+$ diquark mass.
The magnitude of its mass is such that this correlation may be important in
the Fadde'ev equation studies of the strange octet-baryons.
No $0^-$ solution with a mass less than $2~{\rm GeV}$ was found.  This is in
accord with our finding that in the $0^+$ meson channel, there was
insufficient attraction for a clear bound state.

One observes that the diquark mass splitting $M_{us}-M_{ud} = 145$~MeV.  This
may be compared with $m_\Sigma - m_p \approx 250$~MeV.  One might infer from
this that Fadde'ev equation studies, such as the ones described above, may
yield the correct ordering and level-separation of the octet baryons.

\subsection{Vector and Axial-vector Diquarks}
\subsubsection{$f_1=u/d=f_2$}
In ladder approximation the homogeneous Bethe-Salpeter equation for vector
and axial-vector colour-antitriplet diquark correlations has the same form as
Eq.~(\ref{bsqqC}) except that the Bethe-Salpeter amplitude carries a Lorentz
index.  It can be recast into the form of Eq.~(\ref{bsqq}) in the same
manner.

For the axial-vector ($1^+$) diquark channel, the auxiliary amplitude
$\Gamma_{5\,\mu}^{\bar 3\,C}(p,P)$ is identical in form to the vector meson
amplitude in Eq.~(\ref{vamp}).  The calculated mass is listed in
Table~\ref{massdq} and the eigen-vector in Eq.~(\ref{avdqev}).  The
axial-vector diquark mass is larger than that of the vector meson, in
agreement with the argument of Ref.~\cite{CRP87}.  However, it is comparable
to the predicted scalar diquark mass.  Hence the $1^+$ diquark-pole is likely
to provide a contribution to the truncated quark-quark ${\cal T}$-matrix,
employed in the type of simplified nucleon Fadde'ev equation studies
described above, that is comparable to that of the scalar diquark.

In the $1^-$ channel, the auxiliary amplitude $\Gamma_{\mu}^{\bar 3\,C}(p,P)$
has the axial-vector meson form given in Eq.~(\ref{avamp}).  The calculated
vector diquark mass is listed in Table~\ref{massdq} and the eigen-vector in
Eq.~(\ref{avdqev}).  It is too massive to be of importance.

Sub-leading Dirac amplitudes contribute little in these channels. 

\subsubsection{$f_1=u/d$, $f_2=s$}
The auxiliary amplitude for the $1^+$ diquark has the same form as the vector
meson amplitude in Eq.~(\ref{ksamp}), while that for the $1^-$ diquark is
simply $\gamma_5$ times this.  The calculated masses are given in
Table~\ref{massdq} and the eigen-vectors in Eq.~(\ref{avusdq}).  

These results suggest that the axial-vector diquark can be important in
Fadde'ev equation studies of strange baryons, whereas the vector diquark can
be neglected.  Again, sub-leading Dirac amplitudes contribute little in these
channels.

\subsubsection{$f_1=s=f_2$}
The auxiliary amplitude for the $1^+$ diquark is identical in form to that
for the $\phi$ meson while that for the $1^-$ diquark has the form of the 
axial counterpart ($f_1(1510)$), both of which are described in Appendix 
\ref{phiapp}.   The calculated masses are
given in Table~\ref{massdq} and the eigen-vectors in Eq.~(\ref{ssvdqev}).

The low mass of the axial-vector diquark suggests that it can be important in
Fadde'ev equation studies of all strangeness carrying baryons, whereas again
the vector diquark can be neglected.  The sub-leading Dirac amplitudes are
unimportant in these channels.

\section{Summary and Conclusions}
\label{secfive}
We have constructed a crude, confining, separable Ansatz for the ladder
kernel of the two-body Bethe-Salpeter equation [BSE] from the
phenomenologically efficacious $u/d$ and $s$ dressed-quark propagators of
Ref.~\cite{BRT96}.  We have emphasised that no connection can be made between
this crude kernel and the solution of the Dyson-Schwinger equation for the
dressed-gluon propagator.

A very good description of the ground-state, $SU_f(3)$,
isovector-pseudoscalar, vector and axial-vector meson spectrum was obtained.
Scalar mesons and the pseudoscalar $\eta-\eta'$ complex were poorly described 
and this is not unexpected in ladder approximation.
Given the crudity of our construction, our results can be interpreted as
demonstrating the reliability and extent of applicability of the
rainbow-ladder approximation to the quark-DSE/meson-BSE complex.

We found that in the isovector-pseudoscalar meson channel the sub-leading
Dirac components of the Bethe-Salpeter amplitude; i.e., those terms whose
Dirac matrix structure is more than just $\gamma_5$, provide
quantitatively important contributions to the mass ($\sim$ 15\% effects) and 
weak
decay constant ($\sim$ 35\% effects).  These terms are unimportant in the vector
and axial-vector meson channels.

We saw that separable Ans\"atze have a number of shortcomings.  In the
pseudoscalar channel one finds that the $\gamma_5$ and $\gamma_5\gamma\cdot
P$ components of the meson Bethe-Salpeter amplitude are characterised by the
same function, $B(p^2)$, which is not true in general.  One also finds that
the dominant components in the Bethe-Salpeter amplitudes of the vector and
axial-vector mesons are characterised by the same functions that characterise
these components of the pseudoscalar mesons, $B(p^2)$.  More sophisticated
studies indicate that the vector meson amplitudes are narrower in momentum
space.  This indicates that the amplitudes we have obtained should be used
with caution in the calculation of, for example, meson-meson scattering
processes.  

The shortcomings notwithstanding, there are areas of study in hadronic physics
for which the presently provided Bethe-Salpeter amplitudes have significantly
greater dynamical justification than that of currently used approximations.
For example,  hadronic coupling constants 
such as $g_{\rho\pi\pi}$~\cite{MT} and $g_{\gamma\pi\rho}$~\cite{T96}
have been reproduced  from the $\bar{q}q$ structure of 
the mesons in terms of a single dominant Dirac covariant if the  
amplitude is allowed some phenomenological freedom.  A more realistic treatment
is facilitated by the present work.

The ladder kernel also has the defect that it is purely attractive in both
the colour-singlet $\bar q$-$q$ and colour-antitriplet $q$-$q$ channels.
This entails that it yields bound colour-antitriplet diquarks.  This is a
peculiarity of ladder approximation.  Measuring ``order'' by the number of
dressed-gluon lines in the Bethe-Salpeter kernel, ladder approximation is the
lowest order kernel.  Repulsive terms appear at every higher order.  It has
been shown~\cite{BRvS96} that in the isovector-pseudoscalar and vector meson
channels, these repulsive terms are cancelled by attractive terms of the same
order.  This explains why ladder approximation is phenomenologically
successful in these channels.  In the colour-antitriplet diquark channel the
algebra of $SU_c(3)$ entails that the repulsive terms are stronger; they are
not completely cancelled and eliminate the diquark bound states~\cite{BRvS96}.

The artificial diquark spectrum we obtain is nevertheless of contemporary
interest because there have been a number of studies of the covariant,
three-body Fadde'ev equation that use the existence of diquark poles in the
quark-quark ${\cal T}$-matrix to reduce this problem to a two-body,
quark-diquark bound-state problem.  Our constrained, separable Ansatz
indicates that such studies of the baryon should include $SU_f(3)$ scalar and
pseudovector diquarks, since these are low in mass, but can neglect
pseudoscalar and vector diquarks.

The diquark results might be used in the following way.  The study of
Ref.~\cite{BRvS96} suggests that, even though colour-antitriplet states are
not bound, one may associate an inverse correlation-length, $M$, with each
channel; the ``bound-state mass'' providing an estimate of this.  One might
then construct a ``pseudo-pole'' representation of the quark-quark ${\cal
T}$-matrix (for example: $\sim \sum_n\,a_n\,[1-\exp(-[P^2+ M_n^2])]/[P^2+
M_n^2]$), which would not entail asymptotic (unconfined) diquark states but
would provide for a simplification of the covariant, three-body Fadde'ev
equation.

Finally, this study shows that in order to directly connect hadron phenomena
with the dressed-gluon propagator, $D_{\mu\nu}(k)$, one must start with a
form of $D_{\mu\nu}(k)$, as in Ref.~\cite{FR96}.  Other approaches, while
they may provide a useful phenomenology, efficacious in that it correlates
many observables via few parameters, can only loosely constrain
$D_{\mu\nu}(k)$ and hence the nature of the quark-quark interaction in the
infrared.

\acknowledgements This work grew from discussions between R. T. Cahill,
C. D. Roberts and P. C. Tandy.  The authors are grateful to the National
Centre for Theoretical Physics at the Australian National University for
hospitality during a visit where part of this work was conducted.  This work
was supported in part by the National Science Foundation under Grant
Nos. PHY91-13117, INT92-15223 and PHY94-14291 and by the US Department of
Energy, Nuclear Physics Division, under contract number W-31-109-ENG-38.
Some of the calculations described herein were carried out using a grant of
computer time and the resources of the National Energy Research Supercomputer
Center.

\appendix
\section{Bethe-Salpeter equation for equal mass quarks}
\label{appA}
Here we present details of the solution of the ladder
approximation to the Bethe-Salpeter equation in the case of equal mass quarks
($f_1 = f_2$) using a separable Ansatz for the kernel.

We begin with Eq.~(\ref{bsem}) subject to Eqs.~(\ref{giaknl}) and (\ref{ggdf})
with $f_1=f_2$ and $\xi = 1/2$:
\begin{equation}
\Gamma(p,P) = - \frac{4}{3} \int \frac{d^4q}{(2\pi)^4}
 \Delta(p - q) \gamma_\mu
S(q +\case{1}{2} P) \Gamma(q,P) S(q - \case{1}{2} P) \gamma_\mu~.
\end{equation}
The general form of the scalar and pseudoscalar meson amplitudes is \cite{LS69}
\begin{eqnarray}
\label{bses}
\lefteqn{\Gamma^{\rm scalar}(q,P)=}\\
 &  & g_I(q^2,P^2,q\cdot P) I   +  \left[g_P(q^2,P^2,q\cdot P) P_\mu 
   + g_u(q^2,P^2,q\cdot P) u_\mu(q) \right] i\gamma_\mu, \nonumber
\end{eqnarray}
and 
\begin{eqnarray} 
\label{bsep}
\lefteqn{\Gamma^{\rm pseud}(q,P)=}\\
 &  & g_5(q^2,P^2,q\cdot P) \gamma_5
   + \left[g_{P5}(q^2,P^2,q\cdot P) P_\mu + 
   g_{u5}(q^2,P^2,q\cdot P) u_\mu(q) \right] i\gamma_\mu \gamma_5,
\nonumber
\end{eqnarray}
where 
\begin{equation}
\label{uvctr}
\begin{array}{lll}
\displaystyle 
u_\mu(p,\hat P) = \frac{p_\mu + p\cdot\hat P\,\hat P_\mu}
		{p^2+ (p\cdot\hat P)^2}~,&
|u(p)| = \sqrt{u(p,\hat P)^2}~, &
\hat u_\mu(p)= \displaystyle \frac{ u_\mu(p,\hat P)}
				{|u(p)|}~,
\end{array}
\end{equation}
with $\hat P_\mu$ $[\hat P^2 = -1]$ the direction-vector associated with
$P_\mu$.

In Feynman-like gauge it follows from the Fierz identity that there is no
piece proportional to $[\gamma\cdot P,\gamma\cdot q]$.  For mesons which are
even (odd) under charge conjugation, $g_I$, $g_u$, $g_5$ and $g_{P5}$ are
even (odd) functions and $g_P$ and $g_{u5}$ odd (even) functions of $q\cdot
P$.

Defining \mbox{$k_\mu = q_\mu + \case{1}{2} P_\mu$} and \mbox{$l_\mu = q_\mu
- \case{1}{2} P_\mu$} one has
\begin{equation}
\begin{array}{lll}
P\cdot u(q) = 0~, & k\cdot u(q) = l\cdot u(q) = 1~, &
k\cdot u(p) = l\cdot u(p) = q\cdot u(p)~.
\end{array}
\end{equation}

Multiplying Eq.~(\ref{bses}) by $I$, $\gamma\cdot P$ or $\gamma\cdot u(p)$
and taking traces one projects out a set of coupled integral equations for
the scalar meson amplitudes.  Defining
\begin{equation}
\begin{array}{lll}
f(p,P) = g_I(p,P)~,&
W(p,P) = i\,M\,g_P(p,P)~, &
U(p,P) = |u(p)|\;g_u(p,P)~,
\end{array}
\end{equation}
where $ i\,M = \sqrt{P^2}~$, these equations take the following simple form:
\begin{eqnarray}
f(p) & = & \frac{16}{3} \int \frac{d^4q}{(2\pi)^4} \Delta(p - q)
	\left[ T_{ff} f(q) + T_{fW} W(q) + T_{fU} U(q) \right], 
				   \nonumber \\
W(p) & = & \frac{8}{3} \int \frac{d^4q}{(2\pi)^4} \Delta(p - q)
	\left[ T_{Wf} f(q) + T_{WW} W(q) + T_{WU} U(q) \right], 
				 \label{fwu}  \\
U(p) & = & \frac{8}{3} \int \frac{d^4q}{(2\pi)^4} \Delta(p - q) \,
	     \hat u(p,\hat P)\cdot\hat u(q,\hat P)\,
	\left[ T_{Uf} f(q) + T_{UW} W(q) + T_{UU} U(q) \right], 
				   \nonumber 
\end{eqnarray} 
where 
\begin{eqnarray}
T_{ff}^{\rm scalar} & = & k\cdot l\, |\sigma_V|^2 - |\sigma_S|^2~, \nonumber \\
T_{WW}^{\rm scalar} & = & 
	\left(k\cdot l + 2 k\cdot\hat P\,l\cdot\hat P\right) |\sigma_V|^2
			 +  |\sigma_S|^2~,\nonumber \\
T_{UU}^{\rm scalar} & = & 
\left(k\cdot l - \frac{2}{u(q,\hat P)^2} \right)  |\sigma_V|^2
			  +  |\sigma_S|^2  \label{teesd}
\end{eqnarray}
and 
\begin{equation}
\begin{array}{ccccl}
T_{fW}^{\rm scalar} & = & T_{Wf}^{\rm scalar}
	 & = & M\,\Im(\sigma_V^* \sigma_S) 
	+ 2i\,q\cdot\hat P\Re(\sigma_V^* \sigma_S),     \nonumber \\ 
T_{fU}^{\rm scalar} & = & T_{Uf}^{\rm scalar}
	& = & \displaystyle  -\,\frac{2}{|u(q)|}\,  \Re(\sigma_V^* \sigma_S), 
							\nonumber \\
T_{WU}^{\rm scalar} & = & T_{UW}^{\rm scalar}
	& = & \displaystyle 2\, i\,\frac{q\cdot\hat P}{|u(q)|}\, |\sigma_V|^2 ,
							\label{teeso}
\end{array}
\end{equation}
with
\begin{equation}
\begin{array}{cc}
\sigma_V^\ast \equiv \sigma_V\left(k^2\right), & 
\sigma_V \equiv \sigma_V\left(l^2\right),
\end{array}
\end{equation}
and similarly for $\sigma_S$.  In these equations
\begin{equation}
\begin{array}{l}
|\sigma_V|^2 \equiv \sigma_V\,\sigma_V^*~,\\
\\
\Re(\sigma_V^* \sigma_S) \equiv 
\case{1}{2}\left(\sigma_V^* \sigma_S + \sigma_V \sigma_S^*\right)~,\\
\\
\Im(\sigma_V^* \sigma_S) \equiv 
\case{1}{2i}\left(\sigma_V^* \sigma_S - \sigma_V \sigma_S^*\right)~.
\end{array}
\end{equation}

In terms of the functions 
\begin{equation}
\begin{array}{lll}
f(p,P) = i g_5(p,P)~, & 
W(p,P)  =  i\,M\,g_{P5}(p,P)~,& 
U(p,P)  = |u(p)|\;g_{u5}(p,P)~,   \label{pscale}
\end{array}
\end{equation}
the equations for the pseudoscalar states have the form in Eq.~(\ref{fwu})
but with the $T$s replaced by
\begin{eqnarray}
T_{ff}^{\rm pseud} & = & k\cdot l\, |\sigma_V|^2 + |\sigma_S|^2 \nonumber \\
T_{WW}^{\rm pseud} & = & 
	\left(k\cdot l + 2 k\cdot\hat P\,l\cdot\hat P\right) |\sigma_V|^2
			 - |\sigma_S|^2\nonumber \\
T_{UU}^{\rm pseud} & = & 
	\left(k\cdot l - \frac{2}{u(q,\hat P)^2} \right)  |\sigma_V|^2
			  - |\sigma_S|^2, \label{teepd}
\end{eqnarray}
and
\begin{equation}
\begin{array}{ccccl}
T_{fW}^{\rm pseud} & = & -T_{Wf}^{\rm pseud} & = &
	- M\,\Re(\sigma_V^* \sigma_S) 
	+ 2\,i\,q\cdot\hat P\,\Im(\sigma_V^* \sigma_S),  \nonumber \\ 
T_{fU}^{\rm pseud} & = & -T_{Uf}^{\rm pseud}
	  & = &  \displaystyle -\, \frac{2}{|u(q)|}\, \Im(\sigma_V^* \sigma_S), 
							\nonumber \\
T_{WU}^{\rm pseud} & = & T_{UW}^{\rm pseud} 
	 & = & \displaystyle 2\, i\,\frac{q\cdot\hat P}{|u(q)|}\, |\sigma_V|^2~.
			  \label{teepo}
\end{array}
\end{equation}

\subsection{Separable Ansatz}

The form of the gluon propagator $\Delta(p -q)$ appearing in Eq.~(\ref{fwu})
is not yet specified.  Introducing the separable form in Eq.~(\ref{sepeqm})
and taking into account the symmetry properties of the functions $f$, $W$ and
$U$ under \mbox{$p\cdot\hat P\rightarrow -p\cdot\hat P$}, which follow from
charge conjugation symmetry, one obtains the following sets of integral
equations for the scalar mesons:
\subsubsection{ Scalar, ${\cal C} = +$ mesons}
\begin{eqnarray} 
f(p) & = & \frac{16}{3} \int \frac{d^4q}{(2\pi)^4}\, G(p^2)\,G(q^2)\,
	\left[ T_{ff} f(q) + T_{fW} W(q) + T_{fU} U(q) \right], 
				   \nonumber \\ 
W(p) & = & \frac{8}{3} \int \frac{d^4q}{(2\pi)^4}\, 
			p\cdot q\,F(p^2)\,F(q^2)\,
	\left[ T_{Wf} f(q) + T_{WW} W(q) + T_{WU} U(q) \right], 
					\label{bsescpl} \\
U(p) & = & \frac{8}{3} \int \frac{d^4q}{(2\pi)^4} \,
		p\cdot q\,F(p^2)\,F(q^2)\,
	\hat u(p,\hat P)\cdot\hat u(q,\hat P)\,
	\left[ T_{Uf} f(q) + T_{UW} W(q) + T_{UU} U(q) \right]\nonumber 
\end{eqnarray}
\subsubsection{ Scalar, ${\cal C} = -$ mesons}
\begin{eqnarray} 
f(p) & = & \frac{16}{3} \int \frac{d^4q}{(2\pi)^4}\, 
		p\cdot q\,F(p^2)\,F(q^2)\,
	\left[ T_{ff} f(q) + T_{fW} W(q) \right], \nonumber \\ 
W(p) & = & \frac{8}{3} \int \frac{d^4q}{(2\pi)^4} \, G(p^2)\,G(q^2)\,
	\left[ T_{Wf} f(q) + T_{WW} W(q) \right], \\
 U(p)  & = & 0.\nonumber
\end{eqnarray}
The $T_{ff}, T_{fW}, \ldots$ are given by Eqs.~(\ref{teesd},\ref{teeso}).
There are similar equations for the pseudoscalar mesons.

\subsection{Form of the solution using the constrained separable Ansatz}
The separable form of the propagator causes the solutions of these equations 
to be proportional to the functions $G$ and $F$.  For the scalar,
${\cal C}=+$ meson, for instance, one finds that the solution is of the form
\begin{equation}
 \label{fwusol}
\begin{array}{rcl}
f(p) & = & \lambda_f G(p^2), \\    
W(p) & = & -i\,\lambda_W p\cdot\hat P F(p^2), \\    
U(p) & = & \lambda_U \frac{1}{|u(p)|} F(p^2)~.
\end{array}
\end{equation}

Substituting the above Ansatz into Eqs.(\ref{bsescpl}) yields a simple matrix
equation of the form
\begin{equation}
\left(\begin{array}{c} 
       \lambda_f \\ \lambda_W \\ \lambda_U \end{array} \right)
 = K(M) \left(\begin{array}{c} 
       \lambda_f \\ \lambda_W \\ \lambda_U \end{array} \right),
				  \label{kofm}
\end{equation}
where $K(M)$ is a $3\times3$ matrix whose elements are two-dimensional
integrals that are completely determined once $\sigma_{\rm V}$ and
$\sigma_{\rm S}$ are specified.  This equation is then solved by adjusting
the meson mass $M$ until one of the eigen-values of $K$ equals one.  This
procedure can be implemented by introducing an eigen-value, $\mu(M)$, on
the left-hand-side of Eq.(\ref{kofm}); solving for $\mu(M)$ and the
eigen-vector at each value of $M$; and repeating the process until one finds
$M$ such that $\mu(M)=1$.  At this point one also has the Bethe-Salpeter
amplitude for the bound state, which is characterised by the multiplet
$\{\lambda_f, \lambda_W, \lambda_U\}$.

Bound states of unequal mass quarks [see Eqs.~(\ref{sepneqm}) and
Sec.~\ref{usmesons}, for example], are not characterised by a
charge-conjugation quantum number, $C$.  In this case the functions $f$, $W$
and $U$ are complex and Eq.~(\ref{fwusol}) generalises to forms such as
\begin{eqnarray}
f(p)  & = & \lambda_{1u} G_{u}(p^2) + \lambda_{1s} G_{s}(p^2) 
	+ q\cdot \hat P \left[ \lambda_{2u} F_{u}(p^2) 
		+ \lambda_{2s}  F_{s}(p^2) \right] + \ldots . \\ \nonumber
\end{eqnarray}
In this case the analogue of the matrix $K(M)$ in Eq.(\ref{kofm}) is, in
general, a $12 \times12$ matrix, which reduces to a $10 \times 10$ matrix
when residual symmetry under $C$ is taken into account.

\section{Bethe-Salpeter Amplitudes in Separable Approximation}

\subsection{Scalar and Pseudoscalar Mesons.}

\subsubsection{$f_1=u/d=f_2$}
The most general form for the solutions of Eq.~(\ref{bspi}) in the scalar and
pseudoscalar channels are:
\begin{eqnarray}
\label{gampseud}
\Gamma^{\rm pseud}_+ (p,P) & =&  G_u(p^2)\,\left[\lambda_f\,I_D -
      i\,\lambda_W \, \gamma\cdot\hat P \right] i\gamma_5,\\
\label{gamscpl}
\Gamma^{\rm scalar}_+ (p,P) &=&  G_u(p^2)\, \lambda_f\,I_D +
  i\, F_u(p^2)\left[-\lambda_W \,p\cdot\hat P\, \gamma\cdot\hat P +
   \lambda_U\,\frac{1}{|u(p)|}\,\gamma\cdot\hat u(p)\right],\\
\Gamma^{\rm pseud}_-(p,P) &=& F_u(p^2)\left[i\,\lambda_f \,p\cdot\hat P  +
  \lambda_W\,p\cdot\hat P \,\gamma\cdot\hat P
	 -\lambda_U\,\frac{1}{|u(p)|}\,\gamma\cdot\hat u(p)\right]i\gamma_5,\\
\label{gamscmn}
\Gamma^{\rm scalar}_-(p,P) & =&  i\,F_u(p^2)\,\lambda_f\, p\cdot\hat P I_D -
    G_u(p^2)\,\lambda_W\, \gamma\cdot\hat P~,
\end{eqnarray}
where $\hat P_\mu$ $[\hat P^2 = -1]$ is the direction-vector associated with
$P_\mu$, and $u_\mu(p,\hat P)$ is defined in Eq.~(\ref{uvctr}).  We note 
the covariants involving $\gamma\cdot\hat u(p)$ may be brought to a more
familiar form through use of the identity
\begin{equation}
\frac{1}{|u(p)|}\,\gamma\cdot\hat u(p) = \gamma\cdot p 
+ p\cdot\hat P \,\gamma\cdot\hat P  .
\end{equation}

Solving Eq.~(\ref{bspi}) yields the following ${\cal C}=\pm$, pseudoscalar
and scalar eigen-vectors:
\begin{equation}
\label{uures}
\begin{array}{lcccc}
J^{PC}          & 0^{-+} & 0^{++} & 0^{--} & 0^{+-} \\
\lambda_f       & 0.61   &  0.67   &  -0.52   & 0.11 \\
\lambda_W       & -0.045 &  -0.0075 &  0.084  & 0.26\\
\lambda_U       & 0.0    &  -0.050 &  0.024  & 0.0 
\end{array}
\end{equation}
which are normalised in accordance with Eq.~(\ref{gnorm}).

\subsubsection{$f_1=u/d$, $f_2=s$}
Solutions for the  pseudoscalar Bethe-Salpeter amplitudes for the 
$\overline{u}$-$s$-mesons are of the form 
\begin{eqnarray}
\lefteqn{\Gamma^{\rm pseud}(p,P)  = } \nonumber \\
 &  &   \left\{ \rule{0mm}{6mm}
\lambda_{1u} G_{u}(p^2) + \lambda_{1s} G_{s}(p^2)
  - p\cdot\hat P \left[ \lambda_{2u}\,F_{u}(p^2) 
	    + \lambda_{2s} \,F_{s}(p^2)\right] \right.  \nonumber \\
 &  & - \left(\lambda_{3u} G_{u}(p^2) + \lambda_{3s} G_{s}(p^2)
 -p\cdot\hat P\left[\lambda_{4u} F_{u}(p^2) + \lambda_{4s} F_{s}(p^2)\right]
		  \right)i \gamma\cdot\hat P  \nonumber \\
 &  & -i \left. \left[ \lambda_{5u}  F_{u}(p^2) + 
			 \lambda_{5s} F_{s}(p^2) \right]
	  \frac{1}{|u(p)|}\,\gamma\cdot\hat u(p) \right\} i\gamma_5~.
\label{ampk}
\end{eqnarray}
The scalar amplitude has the same form but with $i\gamma_5 \to I_D$. 
The pseudoscalar amplitude given here corresponds to the $K^-$.  The $K^+$ 
amplitude is obtained by making the replacement $p_\mu \rightarrow -p_\mu$.

The calculated pseudoscalar eigen-vector is 
\begin{equation}
\label{useigenvec}
\begin{array}{lcccccccccc}
J^P & \lambda_{1u} & \lambda_{1s} & \lambda_{2u} & \lambda_{2s} & \lambda_{3u}
 & \lambda_{3s} & \lambda_{4u} & \lambda_{4s} & \lambda_{5u} & \lambda_{5s} \\
0^- & 263 & 390 & -1.3 & -3.3 & -60 & -97 & 2.9 & 7.0 & 2.8 & 6.7 \\
\end{array}
\end{equation}
where each of the components is to be multiplied by $10^{-3}$ and the
normalisation is in accordance with Eq.~(\ref{gnorm}).  No true scalar
solution is found.

\subsubsection{$\eta$ Meson.} 
The normalisation condition for the $\eta_{\theta_P}$ meson is
\begin{eqnarray}
\lefteqn{ 2  P_{\mu} =  N_c \int \frac{d^4k}{(2\pi)^4} \times} \nonumber \\
 & & 
\!\!\!\!\!\left\{ \case{1}{3} (\cos \theta_P - \surd{2}\sin \theta_P)^2
\left[ {\rm tr} \left( \Gamma_{\eta}(k,-P)
       \partial_\mu^P S_u(k + \case{1}{2} P) \Gamma_{\eta}(k,P)
   S_u(k - \case{1}{2} P) \right) \right. \right. \nonumber \\
 & &  
\hspace{5 mm} \left. +  {\rm tr} \left( \Gamma_{\eta}(k,-P)
	S_u(k + \case{1}{2} P) \Gamma_{\eta}(k,P)
   \partial_\mu^P S_u(k - \case{1}{2} P) \right) \right] \nonumber \\
 & \!\!+ &\!\!\! \case{1}{3} (\surd{2}\cos \theta_P + \sin \theta_P)^2
	  \left[  {\rm tr} \left( \Gamma_{\eta}(k,-P)
       \partial_\mu^P S_s(k + \case{1}{2} P) \Gamma_{\eta}(k,P)
   S_s(k - \case{1}{2} P) \right) \right. \nonumber \\
 & &  \hspace{5 mm} \left.\left. +
  {\rm tr} \left( \Gamma_{\eta}(k,-P)
	S_s(k + \case{1}{2} P) \Gamma_{\eta}(k,P)
   \partial_\mu^P S_s(k - \case{1}{2} P) \right) \right]\right\}. \label{enorm}
\end{eqnarray}

The formula for the decay constant of the $\eta_{\theta_P}$ meson is:
\begin{eqnarray}
\lefteqn{P^2 f_\eta = \frac{N_c}{\surd 2} \int \frac{d^4k}{(2\pi)^4} \times }
       \nonumber \\
  & & \left\{ \case{1}{3}(\cos \theta_P - \surd{2}\sin \theta_P)^2
	{\rm tr} \left[ \gamma\cdot P \gamma_5 
	S_u(p + \case{1}{2} P) \Gamma_{\eta}(p,P) S_u(p - \case{1}{2} P)
\right] \right. \nonumber \\
  & +  & \left. \case{1}{3}(\surd{2}\cos \theta_P + \sin \theta_P)^2 
	{\rm tr} \left[ \gamma\cdot P \gamma_5 
	S_s(p + \case{1}{2} P) \Gamma_{\eta}(p,P) S_s(p - \case{1}{2} P) 
	\right] \right\}.\label{edecay}
\end{eqnarray}

The positive charge parity solution of Eq.~(\ref{ebse}) has the form:
\begin{eqnarray}
\lefteqn{\Gamma_\eta (q,P) = } \nonumber \\ 
& & \left[\lambda_{fu} G_{u}(q^2) + \lambda_{fs} G_{s}(q^2)
  - \left[\lambda_{Wu}  G_{u}(q^2) 
	+ \lambda_{Ws} G_{s}(q^2)\right]i\gamma\cdot\hat P
			    \right] i\,\gamma_5.
\label{etagamma}
\end{eqnarray}

Solving Eq.~(\ref{ebse}) leads to the following values of the the mass, decay
constant and eigen-vector at the listed values of $\theta_P$:
\begin{equation}
\label{etares}
\begin{array}{lcccccc}
\theta_P & 5^\circ & 0^\circ & -5^\circ & -10^\circ 
			& -90^\circ & -95^\circ \\
M_{\eta_{\theta_P}} & 
	0.549 & 0.513 & 0.475 & 0.436 & 0.357 & 0.399\\
f_{\eta_{\theta_P}} & 
	0.114 & 0.111 & 0.108 & 0.105 & 0.100 & 0.102 \\
\lambda_{fu} &
	0.18  & 0.23 & 0.28 & 0.33 & 0.43 & 0.38 \\
\lambda_{fs} &
	0.47  & 0.41 & 0.35 & 0.29 & 0.19 & 0.25 \\
\lambda_{Wu} &
	-0.041&-0.051 & -0.059 & -0.066 & -0.073 & -0.071\\
\lambda_{Ws} &
	-0.11 & -0.092 & -0.074 & -0.058 &-0.032 & -0.045
\end{array}
\end{equation}
The masses and decay constants are given in GeV.  The last two columns
correspond to the $\eta^\prime$ flavour projection.

\subsection{Vector and Axial-vector Mesons.}
\subsubsection{$f_1=u/d=f_2$}
The on-shell constraint of Eq.~(\ref{vatrans}) entails that in constructing
the general form of the vector meson Bethe-Salpeter amplitude we can work
with the following transverse Euclidean covariants:
\begin{equation}
\label{tvcov}
\begin{array}{ccccc}
p_\nu^T, & \gamma_\nu^T, & p_\nu^T \gamma\cdot p, &
p_\nu^T \gamma\cdot P p\cdot P, & 
\gamma_5 \epsilon_{\mu\nu\lambda\rho}\gamma_\mu p_\lambda P_\rho~.
\end{array}
\end{equation}
The general amplitude is a linear combination of these covariants weighted 
by invariant amplitudes ${\cal F}_i(p^2,P^2, p\cdot P)$.  
Employing the separable Ansatz, Eq.~(\ref{sepeqm}), the Bethe-Salpeter
equation, Eq.~(\ref{bsv}), cannot support contributions to $\Gamma_\nu(p,P)$
that are bilinear in $p$.  Hence, at the mass-shell, the produced vector meson
Bethe-Salpeter amplitude is
\begin{equation}
\Gamma_{\nu}^T(p,P) = p_{\nu}^T F_u(p^2)\hat{\lambda}_1  
+ i\gamma_{\nu}^T G_u(p^2)\hat{\lambda}_2
+ i\gamma_5 \epsilon_{\mu \nu \lambda \rho} 
\gamma_{\mu} p_{\lambda} \hat{P}_{\rho} F_u(p^2)\hat{\lambda}_3 ~.  
\label{vamp}
\end{equation}

For the axial-vector meson the transverse Euclidean covariants are 
\begin{equation}
\begin{array}{ccccc}
p_{\nu}^T\gamma_5 p \cdot P, & \gamma_5\gamma_{\nu}^T, & 
p_{\nu}^T\gamma_5\gamma \cdot p, & 
p_{\nu}^T\gamma_5\gamma \cdot P p \cdot P, & 
\epsilon_{\mu \nu \lambda \rho} \gamma_{\mu} p_{\lambda} P_{\rho}~.
\end{array}
\label{tavcov}
\end{equation} 
Again the terms bilinear in $p$ do not contribute and, using the separable
Ansatz, the produced axial-vector Bethe-Salpeter amplitude is
\begin{equation}
\Gamma_{5\nu}^T(p,P) =  
i\gamma_5\gamma_{\nu}^T G_u(p^2)\hat{\lambda}_1
+ i\epsilon_{\mu \nu \lambda \rho} \gamma_{\mu} p_{\lambda} \hat{P}_{\rho} 
F_u(p^2)\hat{\lambda}_2 ~.  \label{avamp}
\end{equation}

As above, the Bethe-Salpeter equation is a matrix eigen-value problem.  We
obtain the solutions  
\begin{equation}
\label{vaeigenvec}
\begin{array}{lccc}
J^{PC}          &  \hat\lambda_1    & \hat\lambda_2 & \hat\lambda_3 \\
1^{--}          & 0.075         & -0.33         & 0.049 \\
1^{++}          & 0.056         & -0.28         & 0.0   
\end{array}
\end{equation}
corresponding to the $\rho/\omega$ and $a_1/f_1$ channels respectively.
The Bethe-Salpeter amplitudes are normalised according to the vector and
axial-vector generalisations of Eq.~(\ref{gnorm}).

\subsubsection{$f_1=u/d, f_2=s$}
Consider the \mbox{$J^{P}=1^{-}$} $K^{*+}$ meson.  As for the $\omega$ meson,
there are $5$ transverse covariants, which can be taken to be those in
Eq.~(\ref{tvcov}) except that the explicit factor of $p \cdot P$ is no longer
necessary because $u$-$\bar s$-states are not eigen-states of the charge
conjugation operator, $C$.  The general amplitude is a linear combination of
these covariants weighted by invariant amplitudes ${\cal F}_i(p^2,P^2, p\cdot
P)$ where odd powers of $p \cdot P$ are allowed for the same reason.  The
separable Ansatz does not support contributions bilinear in $p$ and hence, 
at the mass-shell, the produced $K^{*+}$ amplitude has the form
\begin{eqnarray}
\Gamma^{\nu}_T(p,P) &=& p_{\nu}^T 
\left( F_u(p^2)\hat{\lambda}_{1u}+F_s(p^2)\hat{\lambda}_{1s} \right) 
+ i\gamma_{\nu}^T 
\left( G_u(p^2)\hat{\lambda}_{2u}+G_s(p^2)\hat{\lambda}_{2s} \right) 
\nonumber \\
&+& i\gamma_{\nu}^T p \cdot \hat{P} 
\left( F_u(p^2)\hat{\lambda}_{3u}+F_s(p^2)\hat{\lambda}_{3s} \right) 
+ ip_{\nu}^T \gamma \cdot \hat{P}  
\left( F_u(p^2)\hat{\lambda}_{4u}+F_s(p^2)\hat{\lambda}_{4s} \right)
\nonumber \\  
&+& i\gamma_5 \epsilon_{\mu \nu \lambda \rho} 
\gamma_{\mu} p_{\lambda} \hat{P}_{\rho}  
\left( F_u(p^2)\hat{\lambda}_{5u}+F_s(p^2)\hat{\lambda}_{5s} \right)~.  
\label{ksamp}
\end{eqnarray}
The amplitude for the $K^{*-}$-meson is obtained by reversing the sign of
$p$, under which the kernel is invariant.   The amplitudes for the $J^P=1^+$ 
$K_1$ meson states are simply $\gamma_5$ times the appropriate form of 
Eq.~(\ref{ksamp}).

The calculated eigen-vectors $\hat\lambda_i$ are
\begin{equation}
\label{kslam}
\begin{array}{lcccccc}
 J^{P} &  &  \hat{\lambda}_1 
  & \hat{\lambda}_2 & \hat{\lambda}_3 & \hat{\lambda}_4 & \hat{\lambda}_5\\
 1^- & u & 0.020 & -0.12 & -4.5 \times 10^{-4} & 2.0 \times 10^{-4}  & 0.014 \\
     & s & 0.046 & -0.21 & -6.5 \times 10^{-4} & 3.4 \times 10^{-4}  & 0.026 \\ 
 1^+ & u & 0.16  & -5.4 \times 10^{-3} & -6.7 \times 10^{-3}& 
    -1.8 \times 10^{-3}& 6.8 \times 10^{-4} \\
     & s & 0.34  & 4.0 \times 10^{-3}  &  -1.5 \times 10^{-2}&
      9.9 \times 10^{-3} & 1.9 \times 10^{-3}     
\end{array}
\end{equation}

\subsubsection{$f_1=s=f_2$}
\label{phiapp}
The Bethe-Salpeter amplitude for the \mbox{$J^{PC}=1^{--}$} $\bar{s}s$ state 
[$\phi$] has
the same form as Eq.~(\ref{vamp}) for the $\rho/\omega$
but with $G_u \to G_s$
and $F_u \to F_s$.  That for the \mbox{$J^{PC}=1^{++}$} state [$f_1(1510)$]
is related in a similar way to Eq.~(\ref{avamp}) for the $a_1/f_1$.  
The calculated eigen-vectors are 
\begin{equation}
\label{philam}
\begin{array}{lccc}
J^{PC} & \hat{\lambda}_1 & \hat{\lambda}_2 & \hat{\lambda}_3 \\
1^{--} &        0.049   & -0.35         & 0.030 \\
1^{++} &        0.0044  & -0.18         &    0.0
\end{array}   
\end{equation}

\subsection{Diquark Correlations}
All diquark eigen-vectors are normalised according to 
\mbox{$\sum_i\,\left| \lambda_i \right|^2 = 1$}.
\subsubsection{$f_1=u/d=f_2$}
The calculated eigen-vectors for the scalar ($0^{+}$) and pseudoscalar
($0^{-}$) diquark correlations are
\begin{equation}
\label{dqud}
\begin{array}{lccc}
J^P     &\lambda_f^{\bar 3}         &\lambda_W^{\bar 3}&\lambda_U^{\bar 3}\\
0^+     & 0.96                 & -0.29                 & 0.0 \\
0^-     & 0.15                 &  0.58                 & -0.80
\end{array}
\end{equation}

The calculated eigen-vectors for the axial-vector ($1^+$) and vector ($1^-$)
diquark correlations are
\begin{equation}
\label{avdqev}
\begin{array}{lccc}
J^P     &\hat{\lambda}_1        &\hat{\lambda}_2        &\hat{\lambda}_3\\
1^-     & 0.12                 & -1.46                 & 0.0 \\
1^+     & 0.16                 & -0.98                 & 0.11
\end{array}
\end{equation}

\subsubsection{$f_1=u/d, f_2=s$}
The calculated eigen-vector for the scalar ($0^+$) diquark correlation is
\begin{equation}
\label{usdqev}
\begin{array}{lcccccccccc}
J^P & \lambda_{1u} & \lambda_{1s} & \lambda_{2u} & \lambda_{2s} & \lambda_{3u}
 & \lambda_{3s} & \lambda_{4u} & \lambda_{4s} & \lambda_{5u} & \lambda_{5s} \\
0^+ & 498 & 802 & -10.4 & -33.2 & -165 & -282 & 5.2 & 14.9 & 4.7 & 11
\end{array}
\end{equation}
where each component is to be multiplied by $10^{-3}$.

The calculated eigen-vectors for the $1^+$ and $1^-$ diquark correlations are
\begin{equation}
\label{avusdq}
\begin{array}{lcccccc}
J^P&  &  \hat{\lambda}_1 & 
      \hat{\lambda}_2 & \hat{\lambda}_3 & \hat{\lambda}_4 & \hat{\lambda}_5\\
1^-& u&  168  &  -4.7 & -0.029  &  -7.0 &   4.0  \\
   & s&  445  &  6.3  & 0.21    &  -17.0&   39.6 \\
1^+& u&  16.2   &  -161 & -0.64 &  0.13 &   12.9 \\
   & s&  45.8   &  -288 & -1.05 &  0.21 &   26.8         
\end{array}
\end{equation}
where again each component is to be multiplied by $10^{-3}$.

\subsubsection{$f_1=s=f_2$}
The calculated eigen-vectors for the $1^+$ and $1^-$ diquark correlations are
\begin{equation}
\label{ssvdqev}
\begin{array}{lccc}
J^P     &\hat{\lambda}_1        &\hat{\lambda}_2        &\hat{\lambda}_3\\
1^-     & 0.030                & -1.64                 & 0.0 \\
1^+     & 0.090                & -0.99                 & 0.061
\end{array}
\end{equation}


\begin{table}
\begin{center}
\begin{tabular}{|c|l|l||l|l|} 
   $f_i$      & $\epsilon_S^{f_i}$ & $\hat m_{f_i}$  
			& $a_{f_i}^{\rm calc.}$~GeV$^2$ 
			& $b_{f_i}^{\rm calc}$~GeV$^2$ \\ \hline
   $u/d$      & ~0.482           & ~0.00811 
			& ~0.0413        & ~0.0281 \\ \hline
   $s$        & ~0.580           & ~0.198 
			& ~0.0385        & ~0.0426 \\ 
\end{tabular}
\end{center}
\caption{\label{tabnewparam}
Values of the fitting parameters, $\epsilon_S^{f_i}$ and $\hat m_{f_i}$, used
in constructing the constrained, separable Ansatz; and the values $a_{f_i}$
and $b_{f_i}$, defined in Eqs.~(\protect\ref{aparam}) and
(\protect\ref{bparam}), calculated using them.  The parameters $\epsilon_S^u$
and $\hat m$ are chosen so as to fit $m_\pi=137.5$~MeV and $f_\pi=92.4$~MeV;
the parameters $\epsilon_S$ and $\hat m_s$ so as to fit $m_K= 493.6$~MeV and
$f_K=113$~MeV.  The values of $\hat m_{f_i}$ listed here correspond to
$m_{u/d} = 4.59$~MeV and $m_s = 112$~MeV.  See Eq.(\protect\ref{mssrto}) and
associated text for further details.}
\end{table}
\begin{table}
\begin{center}
\begin{tabular}{|l|l|l|l|}
 & $m_M^{\rm calc.}$~GeV & $m_M^{\rm calc.~Dom}$ & Expt.\\ \hline
	$\pi$ ($0^{-+}$)
 & 0.139 (fit)  & 0.116   & $\pi^\pm(140),\pi^0(135$) \\ \hline
   $f_0/a_0$ ($0^{++}$)
 & 0.715         & 0.743         &  $f_0(980)$/$a_0(982)$ \\\hline
   $0^{+-}$
 & 1.082          & 1.092         & Not Seen   \\\hline
   $0^{--}$
 & 1.319          & 1.299         & Not Seen  \\\hline
	$K$    
 & 0.494 (fit) & 0.412  &$K^\pm(494),K^0(498)$\\ \hline
   $K_0^\ast$ 
 & unbound      &    unbound    & $K_0^\ast(1430)$  \\\hline
   ${\eta }(\theta_P=5^{\rm 0})$
 & 0.549  &0.472 &  $\eta(547)$ \\ 
   $\eta(\theta_P=0^{\rm 0}) $
 & 0.513 & 0.441 &  \\\hline
   $ \omega/\rho$ 
 & 0.736 & 0.755 & $\omega(782)/\rho(770)$ \\ \hline
   $a_1/f_1$
	& 1.34    & 1.37       &$a_1(1260)/f_1(1285)$ \\\hline
   $K^{\ast}$  
 & 0.854  & 0.866        & $K^{*}$(892)  \\  \hline
   $K_1$
 & 1.39   &   1.39      & $K_1(1270)$, $K_1(1400)$ \\  \hline
   $\phi$ ($\bar s s$ $1^-$)
 & 0.950       & 0.957        &$\phi(1020)$ \\ \hline
   $\bar s s$ $1^+$
 & 1.60        & 1.60         &$f_1(1510)$ \\ \hline
\end{tabular}
\end{center}
\caption{\label{tabmasses} Calculated meson masses compared with experimental
values\protect\cite{PDG94}, when known.  The column labelled with the
superscript ``Dom'' means that the quantity was calculated using only the
leading Dirac amplitude; e.g., $\Gamma_\pi(p,P) \propto i\gamma_5\,G_u(p^2)$ 
for the
pseudoscalar; ``unbound'' means that in ladder approximation our
constrained, separable Ansatz does not yield a stable bound state in the
channel under consideration.}
\end{table}
\begin{table}
\begin{center}
\begin{tabular}{|l|l|l|l|} 
 & $f_M^{\rm calc.}$~GeV & $f_M^{\rm calc.~Dom}$~GeV  & Expt.\\\hline
   $\pi$   
 & 0.0924~(fit)  & 0.056   & $\pi^+$(0.0924)     \\\hline
   $K^\pm$
 & 0.113~(fit) & 0.76  &$K^+(0.113)$         \\\hline  
   $\eta(\theta=5^{\rm 0})$
 & 0.114 & 0.086  &0.094$\pm$0.007 or 0.091$\pm$0.006 \\
   $\eta(\theta=0^{\rm 0}) $
 & 0.111 & 0.082
  &  \\\hline
\end{tabular}
\end{center}
\caption{Calculated weak decay constants compared with experimental
values\protect\cite{PDG94}.  The superscript ``Dom'' has the same meaning as in
Table~\protect\ref{tabmasses}.
\label{tabwkdcay}}
\end{table}
\begin{table}
\begin{center}
\begin{tabular}{|c|c|l|l|l|} 
$f_1$ & $f_2$ & $J^P$ & $M$~GeV & $M^{\rm Dom}$~GeV \\ \hline
$u/d$ & $u/d$ & $0^+$ & 0.737     &  0.653      \\\hline
$u/d$ & $u/d$ & $0^-$ & 1.50      &  1.52      \\\hline
$u/d$ & $s$   & $0^+$ & 0.882     &  0.786    \\\hline
$u/d$ & $s$   & $0^-$ & unbound   &  unbound    \\\hline
$u/d$ & $u/d$ & $1^+$ & 0.949     & 0.958      \\\hline
$u/d$ & $u/d$ & $1^-$ & 1.47      & 1.48      \\\hline
$u/d$ & $s$   & $1^+$ & 1.05      & 1.05       \\\hline
$u/d$ & $s$   & $1^-$ & 1.53      & 1.53       \\\hline
$s$   & $s$   & $1^+$ & 1.13      & 1.13      \\\hline
$s$   & $s$   & $1^-$ & 1.64      & 1.64    \\
\end{tabular}
\caption{Calculated diquark effective-masses.  The superscript ``Dom'' has
the same meaning as in Table~\protect\ref{tabmasses}.
\label{massdq}}
\end{center}
\end{table}
\end{document}